\documentclass[useAMS,usenatbib]{mn2e} 
\usepackage{graphics} 
\usepackage{epsfig, color} 
\usepackage{graphicx}
\usepackage{amsmath,ulem}

  \newcommand{\mbh}{M_{\rm
    BH}} \def\ltsima{$\; \buildrel < \over \sim \;$}
\def\simlt{\lower.5ex\hbox{\ltsima}} \def\gtsima{$\; \buildrel >
  \over\sim\;$} \def\simgt{\lower.5ex\hbox{\gtsima}}

 \def\msun{{\,{\rm M}_\odot}}  \def\del#1{{}}

\title[The resolution bias]{The resolution bias: low resolution feedback
  simulations are better at destroying galaxies. }  \author[]{Martin
  A. Bourne$^{1,\star}$, Kastytis Zubovas$^{2}$ and Sergei Nayakshin$^{1}$
  \\ $^{1}$Department of Physics \& Astronomy, University of Leicester,
  Leicester, LE1 7RH, UK\\ $^{2}$Center for Physical Sciences and Technology,
  Savanori\c{u} 231, Vilnius LT-02300, Lithuania\\ $^{\star}$ {E-mail:~} {\rm
    martin.bourne@le.ac.uk} } 

\begin{document} \date{Received} \pagerange{\pageref{firstpage}--\pageref{lastpage}}
\pubyear{2013} \maketitle \label{firstpage} 

\maketitle 

\begin{abstract} 
Feedback from super-massive black holes (SMBHs) is thought to play a key role
in regulating the growth of host galaxies. Cosmological and galaxy formation
simulations using smoothed particle hydrodynamics (SPH), which usually use a
fixed mass for SPH particles, often employ the same sub-grid Active galactic
nuclei (AGN) feedback prescription across a range of resolutions. It is thus
important to ask how the impact of the simulated AGN feedback on a galaxy
changes when only the numerical resolution (the SPH particle mass) changes.
We present a suite of simulations modelling the interaction of an AGN outflow
with the ambient turbulent and clumpy interstellar  medium (ISM) in the inner
part of the host galaxy at a range of mass resolutions. We find that, with
other things being equal, degrading the resolution leads to feedback becoming
more efficient at clearing out all gas in its path. For the simulations
presented here, the difference in the mass of the gas ejected by AGN feedback
varies by more than a factor of ten between our highest and lowest resolution
simulations. This happens because feedback-resistant high density clumps are
washed out at low effective resolutions. We also find that changes in
numerical resolution lead to undesirable artefacts in how the AGN feedback
affects the AGN immediate environment. 
\end{abstract}

\begin{keywords} galaxies: evolution, active, ISM - quasars: general -
  methods: numerical \end{keywords}  

\section{Introduction}

 Feedback from AGN is often invoked in galaxy formation and cosmological
 simulations \citep[e.g.,][]{SpringelEtal05, Schaye10, Dubois12, Schaye14,
   Vogelsberger2014} as well as in semi-analytical models
 \citep[e.g.,][]{BowerEtAl06, Croton06, Fanidakis12} in order to quench star
 formation in galaxies at the high-mass end of the mass function and reproduce
 a number of observational correlations such as the $M_{\rm BH}-\sigma$
 relation \citep{Ferrarese00, Gebhardt00, Tremaine02, KormendyHo13}. The
 general premise in such models is that the AGN provide a source of negative
 feedback, clearing gas from the host galaxy and inhibiting further star
 formation and AGN activity. 

Outflows on kpc scales with velocities $\simgt 1000$ km s$^{-1}$
\citep[e.g.,][]{CanoDiaz12,
  Maiolino12,CiconeEtal14,CiconeEtAl2015,TombesiEtAl15} and momentum fluxes
exceeding the radiative output of the AGN, $\dot{P}_{\rm AGN}=L_{\rm AGN}/c$,
by factors of upto $\sim 30$ \citep{Dunn10, FergulioEtal10, Moe10, Rupke11,
  Sturm11, FG12, FQ12a, GenzelEtal14, TombesiEtAl15} have been
observed and are believed to be driven by AGN. Such observations provide
compelling evidence that AGN can indeed have an impact on the host galaxy,
playing an important role in establishing observed correlations and thus
vindicating the use of AGN feedback in simulations and semi-analytic models
\citep[see also][]{McNamara2007, Fabian2012, KingPounds15}.  

Observations of local ($z\simlt 0.1$) AGN have found that $\sim 40\%$ of
systems host ``ultra-fast'' outflows (UFOs), with velocities of $v\sim 0.1$ c
\citep{TombesiEtal10,Tombesi2010ApJ} at small radii. Typically such outflows
have mass outflow rates $\dot{M}_{\rm out}\sim 0.1\msun$ yr$^{-1}$ and kinetic
energy fluxes $\dot{M}_{\rm out}v^{2}/2\simeq 0.05L_{\rm Edd}$. Models
\citep{King03, king05} show that when these outflows impact upon the ISM, the
wind shock can reach temperatures of order $\sim 10^{10}-10^{11}$ K. When
radiative cooling of the wind is inefficient, it expands adiabatically and has
the potential to drive the high velocity outflows discussed above and clear
out significant fractions of gas from the host galaxy \citep{FQ12a, ZK12a}.  

Despite the success of cosmological simulations in reproducing large scale
observations \citep[e.g.,][]{Schaye10, McCarthy2010, Fabjan2010,
  Planelles2013, Vogelsberger2014, Schaye14}, they are unable to resolve
scales small enough to probe the ``AGN-engine'' and thus provide limited
insight into the exact processes driving AGN feedback, see \cite{Schaye14} and
\cite{Crain2015} for a detailed discussion. Therefore simulations only model
the effects of the feedback on the ISM, as opposed to the feedback mechanism
itself. Typically such models have to be {\it tuned}, that is, free parameters
of the feedback and other prescriptions have to be varied until a reasonable
fit to a set of calibrating observations is found.  

This unfortunate situation is unlikely to be drastically improved any time
soon because the numerical and physical modelling challenges in AGN and star
formation feedback in cosmological simulations are so great. Nevertheless, in
the interests of the field, it is only fair to ask the question: does this
approach create numerical artefacts that may influence predictions of the
simulations {\it in a systematic way}?    

To give an example, consider how the SMBH mass, $\mbh$, can be limited by a
feedback argument. Suppose that our model for SMBH feedback contains a
parameter $\epsilon_{\rm BH}=\epsilon_{\rm f}\epsilon_{\rm r}$ that defines
the fraction of SMBH rest mass energy that goes into the AGN outflow,
$\epsilon_{\rm BH} \mbh c^2$, where $\epsilon_{\rm r}$ is the radiative
efficiency of the black hole and $\epsilon_{\rm f}$ is the efficiency with
which the radiation couples to the surrounding gas. Some of this energy may be
lost in the outflow-ISM interaction, for example to radiation in cooling
shocks or by escaping the galaxy through low density voids (see below), so
effectively only a fraction, $\epsilon_{\rm ISM}$, of the feedback energy
impacts the host galaxy gas. In this scenario, the maximum SMBH mass is then
limited by
\begin{equation}
\epsilon_{\rm ISM} \epsilon_{\rm BH} \mbh c^2 = M_{\rm gas} \sigma^2\;,
\label{msig1}
\end{equation}
where $M_{\rm gas}$ is the mass of the gas in the host galaxy that AGN
feedback needs to remove from the galaxy and $\sigma$ is the 1D velocity
dispersion. From this simple analytical argument the black holes mass should
be determined by the efficiency parameters such that $M_{\rm
  BH}\propto(\epsilon_{\rm ISM}\epsilon_{\rm BH})^{-1}$. A similar conclusion
is found by \citet{Booth2010} who show that $M_{\rm BH}\propto(\epsilon_{\rm
  BH})^{-1}$, where $\epsilon_{\rm BH}$ is a free parameter of their feedback
model.

AGN feedback is often implemented in galaxy formation simulations as a
sub-grid model for which the black hole efficiency parameter, $\epsilon_{\rm
  BH}$, is set by hand. $\epsilon_{\rm BH}$ is often calibrated in order
to reproduced the observed local black hole scaling relations
\citep[e.g.][]{DiMatteo05, SpringelEtal05, Sijacki07, BoothSchaye09}, with
typical values of $\epsilon_{\rm r}=0.1$ and $\epsilon_{\rm
  f}=0.05-0.15$. However, $\epsilon_{\rm ISM}$, which cannot be directly set
by the simulator, is governed by the ISM modelling i.e. details of the
hydrodynamics, radiative cooling and any other sub-grid ISM routines used in
the simulation. This provides an explanation as to why values for
$\epsilon_{\rm f}$ can differ between simulations. As noted in
\citet{BoothSchaye09} their value of $\epsilon_{\rm f}=0.15$ differs from the
  value of $\epsilon_{\rm f}=0.05$ used by \citet{SpringelEtal05} due to
  compensating for differences between sub-grid ISM modelling. This suggests
  that the effective $\epsilon_{\rm ISM}$ is smaller in \citet{BoothSchaye09}
  compared to \citet{SpringelEtal05}.

From the arguments above, when a simulation is compared to observations, a
constraint is obtained not on $\epsilon_{\rm BH}$ directly but on the product
$\epsilon_{\rm ISM} \epsilon_{\rm BH}$. The danger here is that $\epsilon_{\rm
  ISM}$ is dependent on the numerics and hence the value obtained for
$\epsilon_{\rm BH}$ when calibrating simulations against observed black hole
scaling relations  does not actually directly tell us about the AGN physics
\citep[as already discussed by][]{Schaye14}. We note that $\epsilon_f$ is {\it
  never} considered a prediction of the subgrid AGN feedback models and that
the only requirement for self-regulation of the SMBH growth to occur is that
$\epsilon_f$ is non-zero. Further, it is interesting to note that both the
OWLS \citep{Schaye10} and EAGLE \citep{Schaye14} cosmological simulations had
large differences in resolution and subgrid physics, but used the same value
of $\epsilon_{\rm f}=0.15$. This choice did however require an increase in the
temperature increment of particles heated by AGN feedback in higher resolution
simulations \citep{Schaye14, Crain2015}. This parameter is set by hand and
effectively controls the value of $\epsilon_{\rm ISM}$. The intimate
relationship between $\epsilon_f$ and $\epsilon_{ISM}$, evidenced by these
large-scale simulations, shows that it is important to understand any
potential numerical trends in $\epsilon_{ISM}$, for example with resolution,
before drawing conclusions about AGN feedback mechanisms. Investigation of
these mechanisms is a logical next step in galaxy evolution simulations.

In this paper we perform a resolution study in order to better understand how
numerical resolution can affect the coupling between the SMBH feedback and the
ISM.  As in \citep[hereafter BNH14]{bourne14}, to achieve a certain degree of
realism in modelling the clumpy ISM of real galaxies, we impose a turbulent
velocity field upon the initial smooth gas distribution, and allow clumpy
structures to develop before they are hit with the SMBH outflow. We vary SPH
mass resolution over four orders of magnitude, and we also vary the SMBH
feedback implementation and the cooling prescription used in order to minimise
numerical artefacts. Our numerical simulations allow us to test whether there
are numerical trends in $\epsilon_{\rm ISM}$ for a single SMBH feedback
event. Briefly, our main conclusion is that, in the scenario studied, feedback
in low resolution simulations is far more effective at destroying galaxies
than it is in higher resolution simulations. This indicates, at least
qualitatively, that $\epsilon_{\rm ISM}$ is resolution dependent. 

The paper is structured as follows; section 2 outlines the numerical method
and how the simulations are set up, section 3 highlights the results of the
simulations, section 4 discusses the implications of these results, both
physical and computational, and finally in section 5 we summarise the outcome
of this work.  

\section{Simulation Set-up} 

\subsection{Numerical method}

  We implement the SPHS\footnote{Smooth Particle
    Hydrodynamics with a high-order dissipation Switch.} formalism as
  described in \citet{Read10} and \citet{ReadHayfield12}, within a modified
  version of the N-body/hydrodynamical code GADGET-3, an updated version of
  the code presented in \citet{Springel05}. The second-order
  Wendland kernel \citep{Wendland95,DehnenAly12} is employed for both SPH
  calculations (using 100 neighbours), and weighting of the AGN feedback. The
  simulations are run in a static isothermal potential with a mass profile
  which follows

\begin{equation}
M(R) = \frac{M_{\rm a}}{a}R = \frac{2\sigma_{\rm pot}^{2}}{G}R
\end{equation}

where $M_{a}=9.35\times 10^{9}$ M$_{\odot}$ is the mass within a radius of
$a=1$ kpc and $\sigma_{\rm pot}=\sqrt{GM_{a}/2a}\simeq 142 \mbox{ km s$^{-1}$}$ is the one dimensional velocity dispersion of the potential. In
  order to prevent gravitational forces diverging at small radii we apply a
  softening length of $0.1$ pc.

An ideal gas is used for all simulations with gas pressure given
by $P = (\gamma -1)\rho u$, where $\rho$ and $u$ are the density and internal
energy per unit mass of the gas respectively and $\gamma =5/3$ is the
adiabatic index. Note that the mean molecular weight for the gas is calculated
self consistently in our simulations, however for simplicity we assume
$\mu=0.63$ when plotting temperature. In our fiducial runs, for gas
temperatures above $T=10^{4}$ K, we use a modified version of the optically
thin radiative cooling function of \citet{sazonovetal05}, which includes
Bremsstrahlung losses, photoionisation heating, line and recombination
continuum cooling and Compton heating and cooling in the presence of an AGN
radiation field. For comparison we also carry out runs using the same
prescription but neglect the effect of inverse Compton (IC) cooling against
the AGN radiation field. This is in light of recent theoretical predictions
\citep{FQ12a} and observational constraints \citep{bourne13} that suggest UFOs
are always energy conserving and do not cool via IC processes as was
previously believed \citep{King03}. Below $T=10^{4}$ K, cooling is modelled as
in \citet{Mashchenko08}, proceeding through fine structure and metastable
lines of C, N, O, Fe, S and Si. For simplicity, solar metalicity is assumed
for all cooling functions.  

We impose a `dynamic' temperature floor such that gas cannot cool below a
temperature of
\begin{equation} 
\begin{aligned}
 T_{\rm floor} & = \rho^{1/3}\frac{\mu m_{\rm P}G}{\pi k_{\rm B}}\left(N_{\rm ngb}m_{\rm SPH}\right)^{2/3} \\  
   & \simeq 350\left(\frac{\rho}{10^{-22}{\rm g cm}^{-3}}\right)^{1/3}\left(\frac{\mu}{0.63}\right)\left(\frac{m_{\rm sph}}{1600M_{\rm \odot}}\right)^{2/3}K
\end{aligned} 
\label{T_floor} 
\end{equation} 
where $\rho$ and $m_{\rm SPH}$ are the density and mass of an SPH particle,
respectively, and $N_{\rm ngb}=100$ is the typical number of neighbours. Such
a temperature floor manifests itself as a polytropic equation of state with an
effective $\gamma=4/3$ and is used for purely numerical reasons to guarantee
that the Jeans mass is independent of density and Jeans length scales with the
SPH kernel smoothing length. This ensures that gas clouds are able to collapse
while avoiding spurious fragmentation due to resolution
\citep{RobertsonKravtsov2008, schaye08}. This method, or variants upon it are
widely used in galaxy formation and cosmological simulations alike
\citep[e.g.,][]{schaye08, HobbsEtAl13} and thus, despite not being physically
motivated, is an important ingredient in our study if we are to compare to
resolutions similar to those achieved in cosmological simulations.  

SPH particles that have reached the temperature floor and have a density above
$\rho =10^{-22}$ g cm$^{-3}$ are considered star forming. The properties of
the temperature floor ensures star formation follows a Jeans instability
criterion. We employ a probabilistic approach to convert a fraction of this
gas into stars. Similar in fashion to \citet{katz92}, the probability of a SPH
particle being converted into a star particle in a given time step $\Delta t$
is given by 
\begin{equation}
  P=1-{\rm exp}\left({-\epsilon_{\rm SF}\frac{\Delta t}{\tau _{\rm
        ff}}}\right) 
\end{equation} 
where $\epsilon_{\rm SF}=0.1$ is the assumed star formation efficiency and $
\tau _{\rm ff}\sim\sqrt{3\pi /32G\rho}$ is the local free-fall time of the
gas. Newly formed star particles are equal in mass to the SPH
particles and only interact with other particles through gravity. 

\subsection{Initial conditions}

 The simulations presented here follow a similar setup to those presented in
 BNH14. We wish to investigate the impact of AGN outflows on ambient gas in
 the host galaxy under realistic conditions, which should certainly include
 the fact that the ISM is very non-homogeneous, that is, clumpy. To achieve
 that  condition in the controlled environment of an isolated simulation,
 similar to \citet{HobbsEtal11}, a turbulent velocity field is imposed upon
 the gas. A sphere of gas (cut from a relaxed, glass-like configuration) is
 seeded with a turbulent velocity field using the method of
 \citet{DubinskiEtAl95}, as described in \citet{HobbsEtal11}. Assuming a
 Kolmogorov power spectrum with $P_{\rm v}(k)\sim k^{-11/3}$, where $k$ is the
 wavenumber, the gas velocity can be defined as $\vec{v}=\nabla\times\vec{A}$
 where $\vec{A}$ is a vector potential whose power spectrum is described by a
 power-law with a cutoff at $k_{\rm min}\simeq 1/R_{\rm out}$ (where $R_{\rm
   out}$ is the outer radius of the system and defines the largest scale,
 $\lambda_{\rm max}=2\pi/k_{\rm min}$, on which turbulence is driven). The
 velocity field is generated by sampling $\vec{A}$ in Fourier space. At each
 point $\left(k_x, k_y, k_z\right)$ the amplitudes of the components of
 $\vec{A_{\rm k}}$ are drawn from a Rayleigh distribution with a variance
 given by $<|\vec{A_{\rm k}}|^{2}>$ and phase angles distributed uniformly
 between $0$ and $2\pi$ are assigned. The last step is to take the Fourier
 transform of $\vec{v_{\rm k}}=i\vec{k}\times\vec{A_{\rm k}}$ in order to
 obtain the velocity field in real space.  

The desired set up for the gas distribution, on which the AGN feedback acts
upon, consists of a clumpy gaseous shell with a radial range from $0.1$ kpc to
$1$ kpc and a $10^{8}\msun$ black hole at the center.  This is achieved by
first setting up a gas distribution which initially follows a singular
isothermal sphere potential  from $R_{\rm in}=0.1$ kpc to $R_{\rm out}=1$
kpc. The gas mass fraction within this shell is $f_{\rm g}=M_{\rm g}/M_{\rm
  total}=0.16$, giving a total initial gas mass $M_{\rm g}\simeq 1.6\times
10^{9}\msun$. The system, which initially only consists of SPH particles and
the central sink particle, is then allowed to evolve under the action of a
turbulent velocity field for $1$ Myr, resulting in a clumpy gas
distribution. The turbulent velocity is normalised such that the
root-mean-square velocity, $v_{\rm turb}\simeq \sigma \simeq 154$ km s$^{-1}$
and the gas temperature is initially set to $T\simeq 5.6\times 10^{5}$
K, such that the shell is virialised.   

The black hole is modelled as a $10^{8}\msun$ sink particle. During the
relaxation period gas is added to the sink particle if it falls within our
desired inner boundary for the initial condition of $100$ pc. At the end of
the relaxation period the sink particle mass is reset to our desired black
hole mass  of $10^{8}M_{\odot}$ and the accretion radius is set to $10$
pc. This results in particles at small radii with prohibitively small time
steps being removed whilst allowing us to still be able to follow the inflow
of dense filaments to small radii during and after the AGN outburst. However
to prevent the removal of gas directly heated by the AGN feedback, SPH
particles that are not bound to the collective mass of the sink particle and
background potential (within the SPH particles radial position) are not
accreted. Here we present simulations that  initially have $N_{\rm
  SPH}=10^{3},10^{4},10^{5}$ and $10^{6}$ SPH particles. 

\begin{table*} 
\centering 
\begin{tabular}{| c | c | c | c | c |} 
\hline  Run & $N_{\rm SPH}$ & $m_{\rm SPH}$ $(M_{\odot})$ &  $f_{\rm BH}N_{\rm
  ngb}$ & cooling\\  \hline FN3c & $10^{3}$ & $1.6\times 10^{6}$ & $10^{2}$ &
\citet{sazonovetal05} \\  FN4c & $10^{4}$ & $1.6\times 10^{5}$ & $10^{2}$ &
\citet{sazonovetal05} \\ FN5c & $10^{5}$ & $1.6\times 10^{4}$ & $10^{2}$ &
\citet{sazonovetal05} \\ FN6c & $10^{6}$ & $1.6\times 10^{3}$ & $10^{2}$ &
\citet{sazonovetal05} \\ FN3h & $10^{3}$ & $1.6\times 10^{6}$ & $10^{2}$ &
\citet{sazonovetal05}, no Compton cooling \\ FN4h & $10^{4}$ & $1.6\times
10^{5}$ & $10^{2}$ & \citet{sazonovetal05}, no Compton cooling \\ FN5h &
$10^{5}$ & $1.6\times 10^{4}$ & $10^{2}$ & \citet{sazonovetal05}, no ompton
cooling \\ FN6h & $10^{6}$ & $1.6\times 10^{3}$ & $10^{2}$ &
\citet{sazonovetal05}, no Compton cooling \\ & & & & \\ FM3c & $10^{3}$ &
$1.6\times 10^{6}$ & $10$ & \citet{sazonovetal05} \\ FM4c & $10^{4}$ &
$1.6\times 10^{5}$ & $10^{2}$ & \citet{sazonovetal05} \\ FM5c & $10^{5}$ &
$1.6\times 10^{4}$ & $10^{3}$ & \citet{sazonovetal05} \\ FM6c & $10^{6}$ &
$1.6\times 10^{3}$ & $10^{4}$ & \citet{sazonovetal05} \\ FM3h & $10^{3}$ &
$1.6\times 10^{6}$ & $10$ & \citet{sazonovetal05}, no Compton cooling \\ FM4h
& $10^{4}$ & $1.6\times 10^{5}$ & $10^{2}$ & \citet{sazonovetal05}, no Compton
cooling \\ FM5h & $10^{5}$ & $1.6\times 10^{4}$ & $10^{3}$ &
\citet{sazonovetal05}, no Compton cooling \\ FM6h & $10^{6}$ & $1.6\times
10^{3}$ & $10^{4}$ & \citet{sazonovetal05}, no Compton cooling \\ \hline
\end{tabular} 
\caption{Summary of simulations showing (l-r) run name, initial number of SPH
  particles ($N_{\rm SPH}$), mass of a single SPH particle ($m_{\rm SPH}$),
  number of black hole neighbours heated during feedback ($f_{\rm BH}N_{\rm
    ngb}$) and the cooling prescription used. Run nomenclature takes the form
  F{\it XYZ} where {\it X} defines whether the thermal energy of the AGN
  feedback is deposited into a fixed number of neighbours ({\it N}) or a fixed
  mass ({\it M}) at all resolutions, {\it Y}$={\rm log}_{10}(N_{\rm SPH})$ and
  {\it Z} defines runs in which cooling due to IC processes is ({\it c}) and
  is not ({\it h}) included. }
\label{sims} 
\end{table*} 

\subsection{AGN feedback model}

 Even at the resolutions presented in this paper we are unable to directly
 model the feedback mechanism of the AGN, however we can model the effect of
 the feedback on the ISM. Models of UFOs colliding with the ISM have been
 particularly successful in explaining observational correlations
 \citep[e.g.,][]{King03, king05, ZK12a, FQ12a}. In these models the UFO, with
 a velocity $v\sim 0.1c$, shocks against the ISM, driving a reverse wind shock
 and a forward shock in the ISM. The wind shock can reach temperatures of
 $\sim 10^{10}-10^{11}$ K and expand through thermal pressure, driving out
 material of the ISM. As in \citet{costa14}, it is the effect of the reverse
 wind shock that we attempt to mimic in our feedback method. Similar to
 \cite{DiMatteo05} we thermally couple the feedback to neighbouring gas
 particles in a kernel-weighted fashion. During a time step of length $\Delta
 t$, the energy released by the AGN is given by 
\begin{equation}
E_{\rm therm}=\epsilon_{\rm f} L_{\rm AGN}\Delta t 
\end{equation} 
where $\epsilon_{\rm f}=0.05$ is the efficiency with which the AGN luminosity
couples to the ambient gas, as defined in the introduction and $L_{\rm AGN}$
is the AGN luminosity. Our chosen value for $\epsilon_{\rm f}$ is physically
motivated by models of UFOs which are expected to have a kinetic luminosity
$\dot{E}_{kin, UFO}=(\epsilon_{\rm r}/2)L_{\rm AGN}\simeq 0.05L_{\rm AGN}$
\citep[e.g.,][]{king05, ZK12a}. For simplicity we set the AGN duration to 1
Myr and $L_{\rm AGN}$ to the Eddington luminosity, 
\begin{equation}
 L_{\rm Edd} = \frac{4\pi GM_{\rm BH}c}{\kappa}
\end{equation} 
where $M_{\rm BH}=10^{8}M_{\odot}$ is the black hole mass and $\kappa
  =\sigma_{\rm T}/m_{\rm p}$ is the electron scattering opacity (where
  $\sigma_{\rm T}$ is the Thompson cross-section and $m_{\rm p}$ is the proton
  rest mass) and G is the gravitational constant. The energy given to an SPH
particle, $E_{\rm inj}$, is then given by
\begin{equation} 
E_{\rm inj, k} = E_{\rm therm}\frac{m_{\rm SPH} W(r_{\rm k}-r_{\rm BH}, h_{\rm
    bh})}{\rho_{\rm g}(r_{\rm BH})} \;,
\end{equation}
where $m_{\rm SPH}$ is the mass of an SPH particle, $W(r_{\rm k}-r_{\rm BH},
h_{\rm BH})$ is the kernel weight of the SPH particle relative to the black
hole, $h_{\rm BH}$ is the black hole {\it smoothing} length, calculated over
$f_{\rm BH}N_{ngb}$ neighbours (see table \ref{sims}) and $\rho_{\rm g}(r_{\rm
  bh})$ is the gas density at the location of the black hole. This approach
ensures that gas closer to the black hole is heated to a higher temperature
than gas further away. The total mass heated per time step is given by  
\begin{equation} 
M_{\rm heat}\simeq f_{\rm BH}N_{\rm ngb}m_{\rm SPH}\;,
\end{equation}
where $f_{\rm BH}$ is the ratio of the number of black hole neighbours be
  heated. We consider two main scenarios, one in which $f_{\rm BH}=1$ at all
resolutions and one in which we approximately heat a fixed mass at each
resolution and so set $f_{\rm BH}=N_{\rm SPH}/10^{4}$. This choice of $f_{\rm
  BH}$ is a balance between heating a sufficient number of particles in the
lowest resolution simulations and not heating an excessive number of particles
at high resolution.  

\subsection{Summary of simulations}

 A summary of the simulations is given in table \ref{sims}. We use a
 nomenclature of the form FNYZ or FMYZ, where  ``FN'' signifies that a fixed
 number of SMBH SPH particle neighbours are heated by the feedback
 independently of the SPH particle number used in the simulation. This means
 that  $f_{\rm BH} =1$ for such simulations. ``FM'', on the other hand, stands
 for a fixed mass of SMBH neighbour particles being heated. In these runs the
 number of SPH particle neighbours over which the SMBH feedback is spread
 depends on the numerical resolution of the simulation, and we set $f_{\rm
   BH}=N_{\rm SPH}/10^{4}$, at all resolutions. The number Y$=
 \log_{10}(N_{\rm SPH})$ encodes the total number of SPH particles
 used. Finally, {\it Z} is either ``h'' or ``c'', and marks runs in which
 cooling\footnote{Compton heating due to the AGN radiation field is included
   for gas with $T\leq 1.9\times 10^{7}$ K in all simulations.} due to IC
 processes is included (c) or not (h). 

\section{Results}

\subsection{Pre-feedback properties of the ISM}

\begin{figure} 
\psfig{file=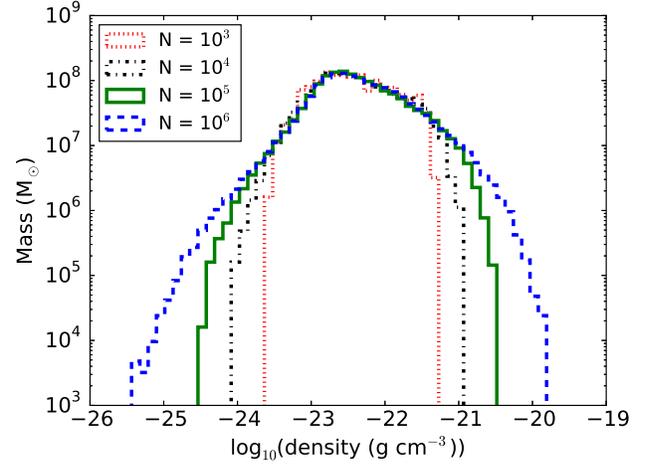,width=0.5\textwidth,angle=0}
\caption{Gas density distribution at 1 Myr for simulations with $10^{3}$
  (red), $10^{4}$ (black), $10^{5}$ (green) and $10^{6}$ (blue)
  particles. Both the high and the low density tails of the gas density
  distribution are better resolved as the mass resolution of the simulation is
  improved.} \label{rho_dist} \end{figure}

Before investigating how the feedback interacts with the ISM we compare the
properties of the ISM itself at different resolutions. Figure \ref{rho_dist}
shows the density distribution for the gas, at different resolutions, after
1~Myr i.e. just before the feedback turns on. At this point in the simulation
the gas distribution is identical for all of the runs at the same resolution,
e.g. the blue curve in the figure is the same for the runs FN6c, FN6h, FM6c
and FM6h. Figure \ref{rho_dist} shows that the lowest resolution runs, with
$10^{3}$ particles, probe a much narrower density range than the runs in which
$10^{6}$ particles are used. The highest resolution runs thus resolve the
density distribution tails at both the low and high density ends. This means
that with improved resolution we are able to better distinguish the high and
low density phases of the ISM, which, as we show below, can have a large
impact on the efficiency of AGN feedback.  

\begin{figure} 
\psfig{file=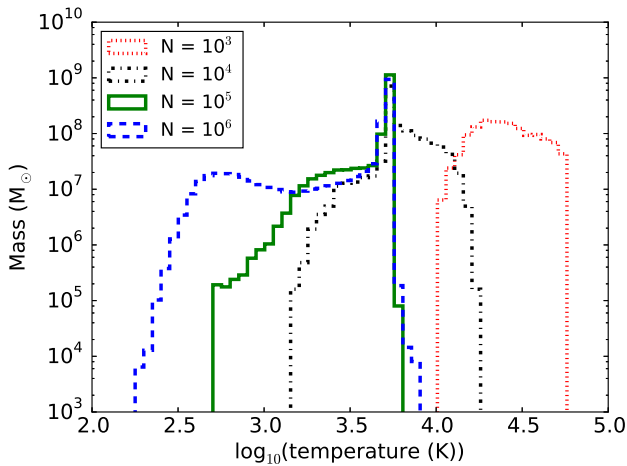,width=0.5\textwidth,angle=0}
\caption{Gas temperature distribution at 1 Myr for simulations with $10^{3}$
  (red), $10^{4}$ (black), $10^{5}$ (green) and $10^{6}$ (blue) particles. The
  low-temperature (dense) gas is completely unresolved in the low resolution
  simulations.} 
\label{T_dist}
\end{figure}

 Figure \ref{T_dist} shows the temperature distribution for the gas, at
 different resolutions, after 1~Myr. In this figure we can see that in the
 lowest resolution simulations ($10^{3}$ particles) the gas remains warm,
 $T>10^4$ K. This is due to the dynamical temperature floor that we employ
 (which is a common approach in cosmological and galaxy formation simulations,
 see text after equation \ref{T_floor}). As the mass resolution is increased the
 gas can cool to progressively lower temperatures. It can also be seen that in
 simulations with $10^{5}$ and $10^{6}$ particles the maximum temperature of
 the gas (before any AGN feedback is initiated) converges to $T\sim 10^{3.7}$
 K. The reason for this is that at lower temperatures ($T \simlt 10^{4}$ K)
 the gas cools much slower, so that there tends to be a lot of gas ``piling
 up'' at $T \sim 10^4$ K. 

\subsection{Overview of numerical resolution trends}
 
\begin{figure*}
\centerline{\psfig{file=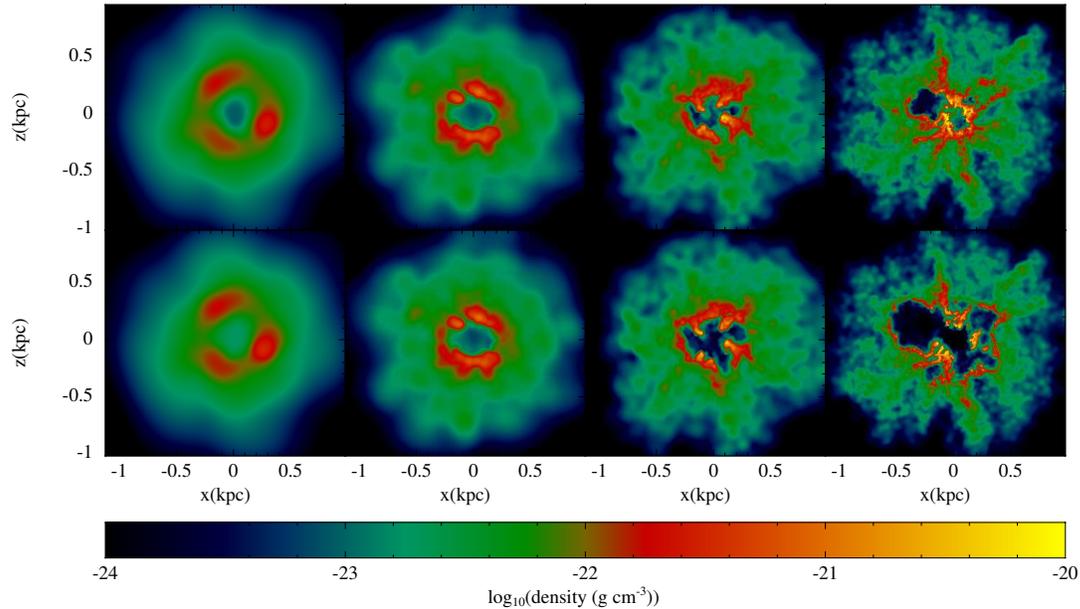,width=0.8\textwidth,clip=true}}
\caption{Density slices at $y=0$ through simulation domains at 1.5 Myr. The
  top row shows runs FM3h, FM4h, FM5h and FM6h from left to right respectively
  while the FN3h, FN4h, FN5h and FN6h runs are shown from left to right
  respectively on the bottom row. }
\label{cross_rho} 
\end{figure*}
\begin{figure*}
\centerline{\psfig{file=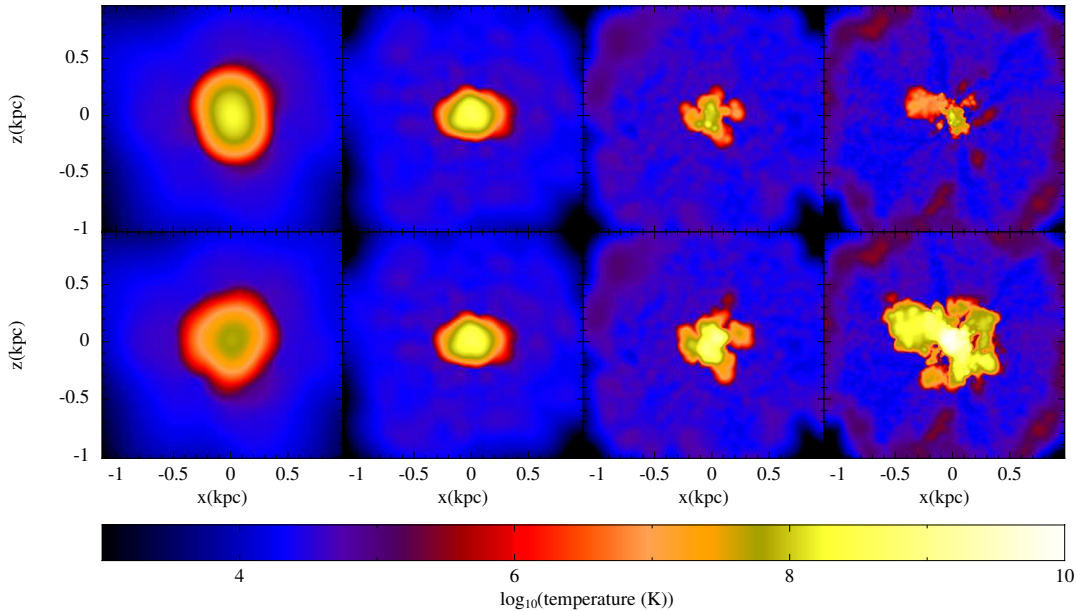,width=0.8\textwidth, clip=true} }
\caption{Temperature slices at $y=0$ through  simulation domain showing gas
  temperature at 1.5 Myr. The top row shows runs FM3h, FM4h, FM5h and FM6h
  from left to right respectively while the FN3h, FN4h, FN5h and FN6h runs are
  shown from left to right respectively on the bottom row. }
\label{cross_T} 
\end{figure*}

 Figure \ref{cross_rho} shows rendered images of gas density slices through
 the simulation domain at $y=0$ after $1.5$ Myr. The top row shows the FM3h,
 FM4h, FM5h and FM6h runs from left to right respectively while the bottom row
 shows the FN3h, FN4h, FN5h and FN6h runs from left to right respectively. The
 figure clearly illustrates the increasing complexity of structure that can be
 resolved with improved resolution. In the low resolution runs (left panels)
 there is a fairly symmetrical swept-up shell of high density gas, while the
 high resolution runs (right panels) consist of compressed high density
 filaments and cleared low density channels through which hot gas can escape.
 
Figure \ref{cross_T} shows the respective temperature slices. At low
resolution (left panels) the feedback outflow sweeps up essentially everything
in its path, with no cold gas left at small radii, whilst the hot gas is
contained in the central regions only. In contrast, in the higher resolution
runs (right panels) the cold gas is still present in clumps and filaments at
small radii, whereas the heated gas escapes through low density channels and
is now more spatially extended. Since it is likely that the cold gas is the
source of efficient SMBH growth, these results show that not only SMBH
feedback but also SMBH growth is affected by the numerical resolution
artefacts i.e. at low resolution there is a lack of high density cold gas
clumps.

The simulations also show stark differences in gas thermodynamical properties
between the runs in which the feedback is coupled to a fixed mass (the FM
series of runs) versus those with a fixed neighbour number (the FN series of
simulations). For instance, due to the differences in the feedback
implementation, in FN6h (bottom right panels of figures \ref{cross_rho} and
\ref{cross_T}) a factor $\sim100$ times less mass is heated than in FM6h (top
right panels of figures \ref{cross_rho} and \ref{cross_T}) and hence the gas
is heated to much higher temperatures in FN6h than in FM6h. This results in
the feedback in FN6h being more effective at clearing gas from the central
region. However there is still cold, dense, in-flowing gas present at small
radii.  

\subsection{Impact of feedback on the ISM}

\subsubsection{Resolving the ISM density structure}

\begin{figure}
\centerline{\psfig{file=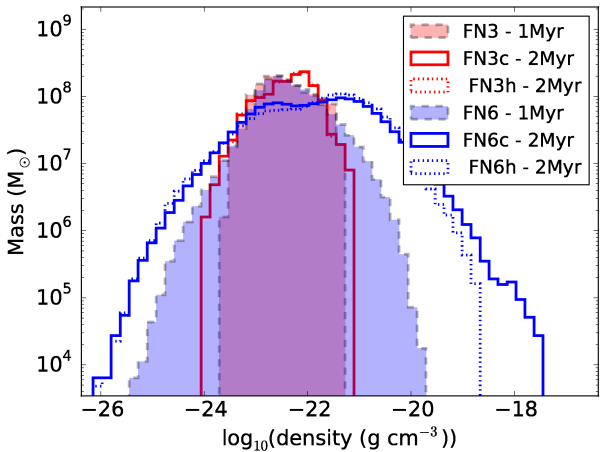,width=0.5\textwidth,angle=0}}
\caption{Comparison of gas density distributions at 1 Myr (filled) and 2 Myr
  (not filled) for FN3 (red) and FN6 (blue) runs with and without IC cooling
  (solid and dotted respectively). It is clear to see that in the FN3 runs the
  density distribution changes very little, while in the FN6 runs the gas can
  be compressed to considerably higher densities.} 
\label{rho_dist_comp_ngb} 
\end{figure} 

\begin{figure}
\psfig{file=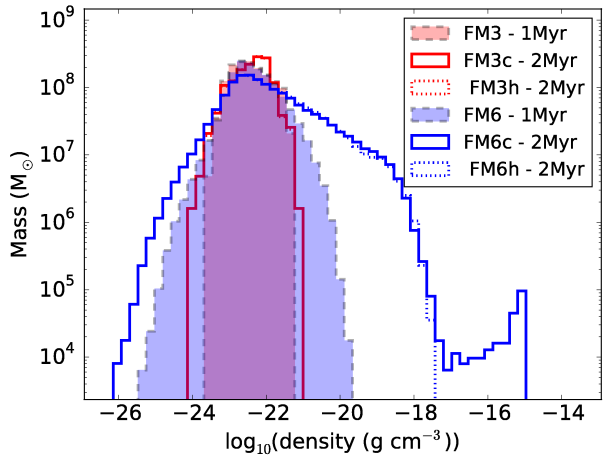,width=0.5\textwidth,angle=0}
\caption {As in figure \ref{rho_dist_comp_ngb} except for the FM3 and FM6
  runs. However, the FM6c run exhibits a particularly high density feature not
  seen in other runs.}
\label{rho_dist_comp_mass} 
\end{figure}

 Figure \ref{rho_dist_comp_ngb} compares the density distribution at 1Myr
 (filled) and 2Myr (not filled) (i.e. before and after the AGN outburst) for
 the FN6 (blue) and FN3 (red) runs. In the FN3 runs there is very little
 evolution in the density distribution with time. However, in the FN6 runs the
 highest value of gas density reached in the simulation increases by
 approximately two orders of magnitude, especially when IC cooling is included
 (compare the dotted and solid lines).   

Similar behaviour is seen when comparing the FM3 and FM6 runs in figure
\ref{rho_dist_comp_mass}. The FM6c run exhibits a particularly high density
feature not seen in other runs. While the mass in this feature corresponds to
only $\sim$ one resolution element, it is interesting to note its
existence. The likely cause of this clump is two-fold. Compared with the FM6h
run, the gas in the FM6c run can cool via IC processes allowing it to collapse
to higher densities. Furthermore, comparing with the FN6 runs, the gas heated
directly by the AGN feedback does not reach such high temperatures in the FM6
runs, potentially resulting in cooler clumps that can reach higher densities. 

In the higher resolution runs the AGN feedback is able to compress gas to much
higher densities, which could result in triggered star formation
\citep[e.g.][]{Elbaz09, Gaibler12, NZ12, Silk13, ZubovasEtal13b,
  Cresci15}. The figure demonstrates that compression of the ISM into high
density features is largely missed in the low resolution runs probably because
the ISM structure is under-resolved.

\subsubsection{Resolving outflows and inflows}

\begin{figure} 
\psfig{file=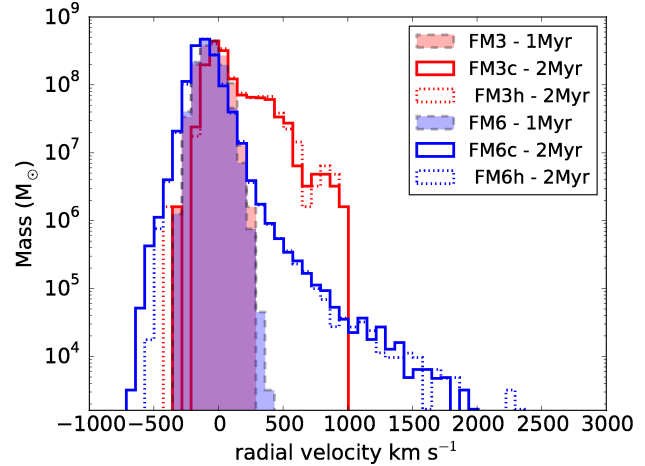,width=0.5\textwidth,angle=0}
\caption{Comparison of radial velocity distributions at 1 Myr (filled) and 2
  Myr (not filled) for FM3 (red) and FM6 (blue) runs with and without IC
  cooling (solid and dotted respectively). Both high and low resolution runs
  produce gas out-flowing with velocities of order $1000$ km s$^{-1}$, however
  the high resolution runs (FM6) show far stronger gas inflows than the low
  resolution runs (FM3).} 
\label{vr_dist_comp_mass} 
\end{figure}
 
\begin{figure} 
\psfig{file=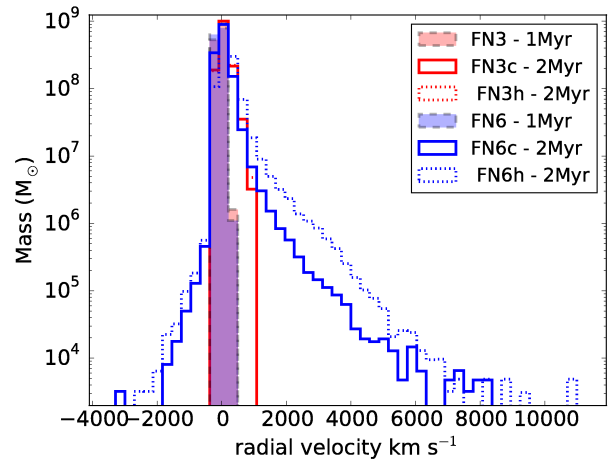,width=0.5\textwidth,angle=0}
\caption{As in figure \ref{vr_dist_comp_mass} except for the FN3 and FN6
  runs. Due to the significantly higher temperatures reached in the FN6 runs,
  compared to the FM6 runs, the gas outflows can reach much higher
  velocities.}
 \label{vr_dist_comp_ngb} 
\end{figure}

  The radial velocity of gas in the simulations is also affected by numerical
  resolution. Contrasting the FM3 and FM6 runs, figure \ref{vr_dist_comp_mass}
  shows that whilst both high and low resolution runs produce gas out-flowing
  with velocities of order $1000$ km s$^{-1}$, the same cannot be said about
  the in-flowing gas: the high resolution runs (FM6) show far stronger gas
  inflows than the low resolution runs (FM3).   The same behaviour is found
  when comparing the FN3 and FN6 runs in figure \ref{vr_dist_comp_ngb}. It is
  interesting to note that for this implementation of feedback, the
  out-flowing gas can reach much higher velocities in the FN6 run than in the
  FN3 run (by a factor of $\sim 10$). This can be attributed to the much
  higher temperatures achieved in the FN6 run. The physical reason for the
  in-flowing gas present only in the high resolution runs is the previously
  emphasised inability of the low resolution simulations to model the high
  density features properly. The high density clumps and filaments present at
  high resolution are artificially smoothed in lower resolution runs. This
  results in the high density gas being far less resilient to feedback in the
  low resolution runs and hence being blown away with the rest of the
  gas. Needless to say, this is a serious artefact as the SMBH may be fed by
  exactly this high density gas falling into the centre of the galaxy despite
  the SMBH feedback \citep[e.g.,][]{Nayakshin14}.

\subsection{Efficiency of feedback versus numerics}

\subsubsection{The over-cooling problem}

Supernova feedback simulations show a well known ``over-cooling problem''
which affects simulation results and is discussed at length in
\citet{DVS12}. We, like other studies \citep[e.g.][]{BoothSchaye09,
  McCarthy2010, McCarthy11,LeBrun14, Schaye14, Crain2015}, find that this
  problem also exists in AGN feedback simulations. The maximum temperature of
the gas directly heated by the feedback, that is the SPH neighbours of the
SMBH particle in which the feedback energy is directly deposited, is inversely
proportional to the total mass of the gas heated. The radiative cooling rate
of the gas is a strong function of temperature in certain temperature
ranges. Therefore, the impact of radiative cooling on the thermal evolution of
this gas depends in a complicated fashion on the number or total mass of SPH
particles in which the feedback energy is injected. In low resolution
simulations it is likely that the injected energy is spread over an
unrealistically large mass of ambient gas. This typically means that this
feedback-heated gas cools on timescales much shorter than one would physically
expect. 

\begin{figure} 
\psfig{file=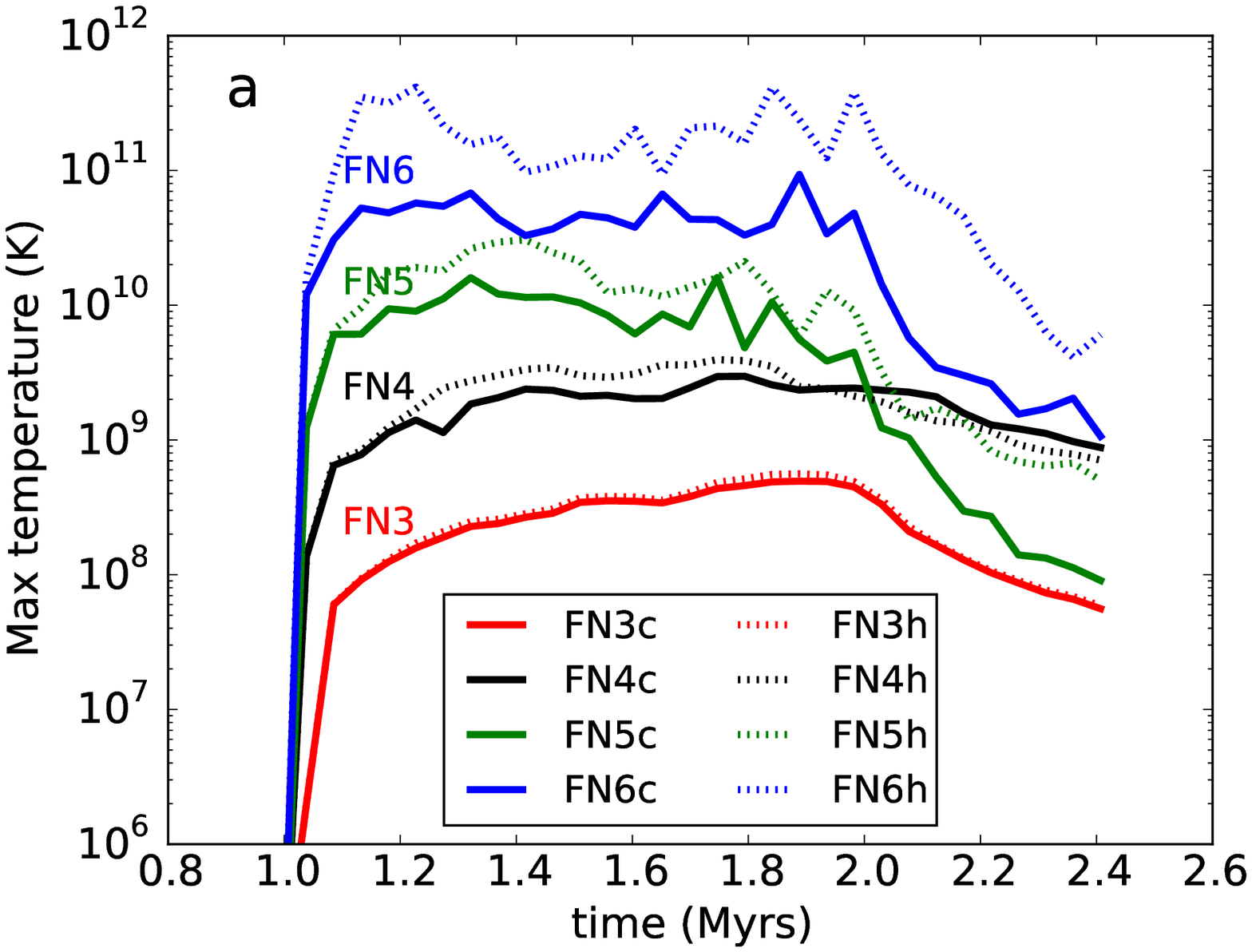,width=0.5\textwidth,angle=0}
\psfig{file=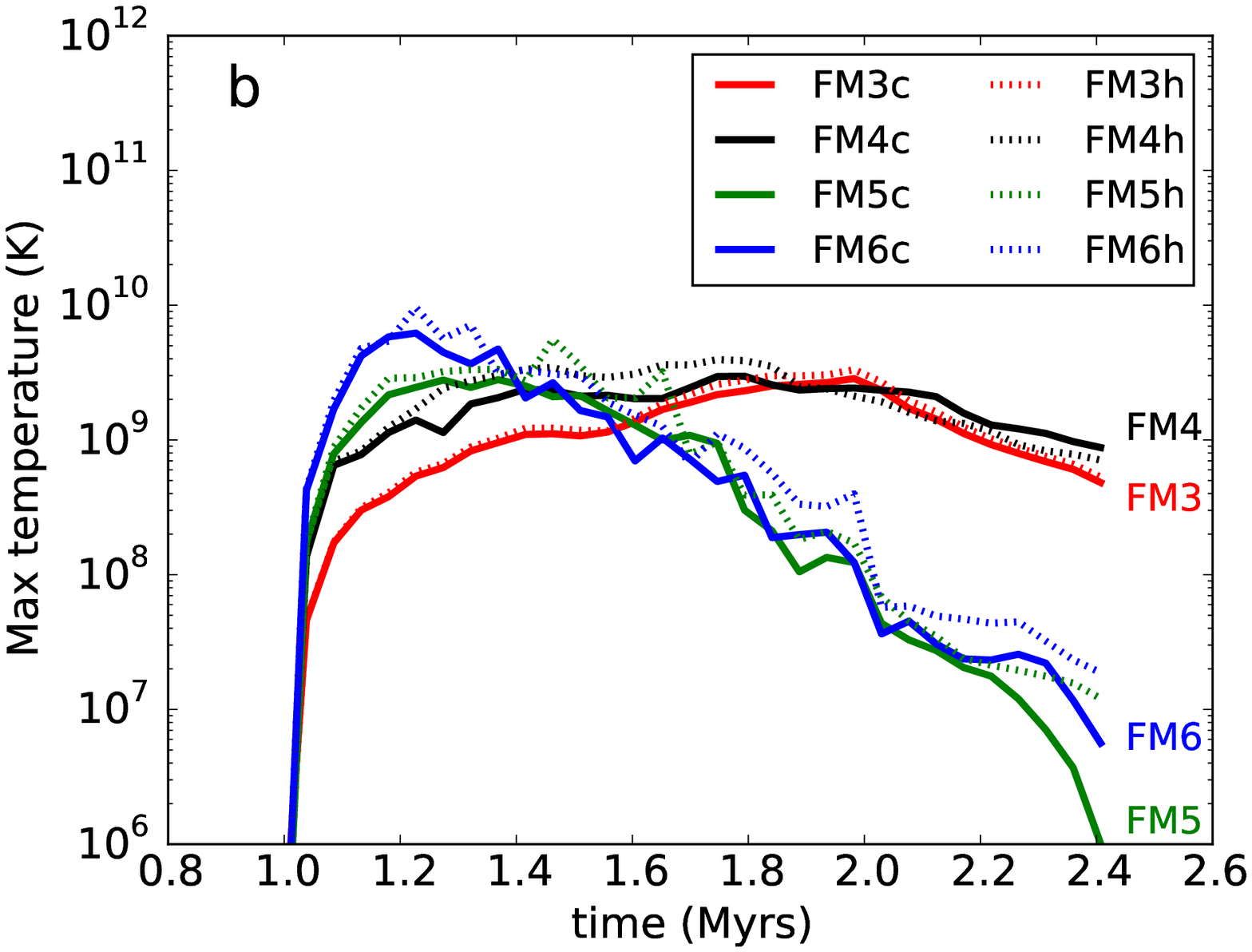,width=0.5\textwidth,angle=0}
\caption{Time evolution of the maximum gas temperature. The top panel (a)
  shows the FN3 (red), FN4 (black), FN5 (green) and FN6 (blue) runs whilst the
  bottom panel (b) shows the FM3 (red), FM4 (black), FM5 (green) and FM6
  (blue) runs. Solid and dotted lines indicate runs with and without IC
  cooling.This figure shows that the temperature to which gas is heated to is
  strongly dependent on the mass of gas heated.}
 \label{Tmax} 
\end{figure}

 Figure \ref{Tmax} shows the time evolution of the instantaneous maximum gas
 temperature for simulations in which the feedback energy is injected into a
 fixed number of SPH particles ($\sim 100$) during each time step (FN3 (red),
 FN4 (black), FN5 (green) and FN6 (blue)) in the top panel (a) and for
 simulations in which the feedback energy is injected into a fixed mass ($\sim
 1.6\times 10^{7} M_{\odot}$) of gas during each time step (FM3 (red), FM4
 (black), FM5 (green) and FM6 (blue)) in the bottom panel (b). The solid and
 dotted lines show runs with and without IC cooling respectively. Considering
 first the fixed number of neighbours (FN) runs in the top panel (a), it
 roughly follows that each order of magnitude improvement in resolution
 results in an order of magnitude increase in temperature. This is because the
 feedback energy is injected into a factor $\sim 10$ times less gas mass for
 each factor $10$ increase in mass resolution.
   
    The fixed mass (FM) runs in the bottom panel (b) lead to more consistent
    results in that the maximum temperature of gas varies much less between
    the different resolution simulations, as is expected. However, at later
    times the higher resolution runs FM5 and FM6 do differ significantly from
    the lower resolution curves. We believe this is caused by differences in
    gas properties beyond the immediate feedback deposition region.  As we
    know from figure 3, there is more dense gas near the black hole in the
    better resolved simulations. This denser gas is hence better at radiating
    the feedback energy away than in the lower resolution  runs.
   
    Finally we note that if included, IC cooling dominates the cooling
    function at high temperatures for gas close to the SMBH. This explains why
    the dotted and dashed curves in figure \ref{Tmax} only exhibit differences
    when gas is heated to high temperatures.

\begin{figure} 
\psfig{file=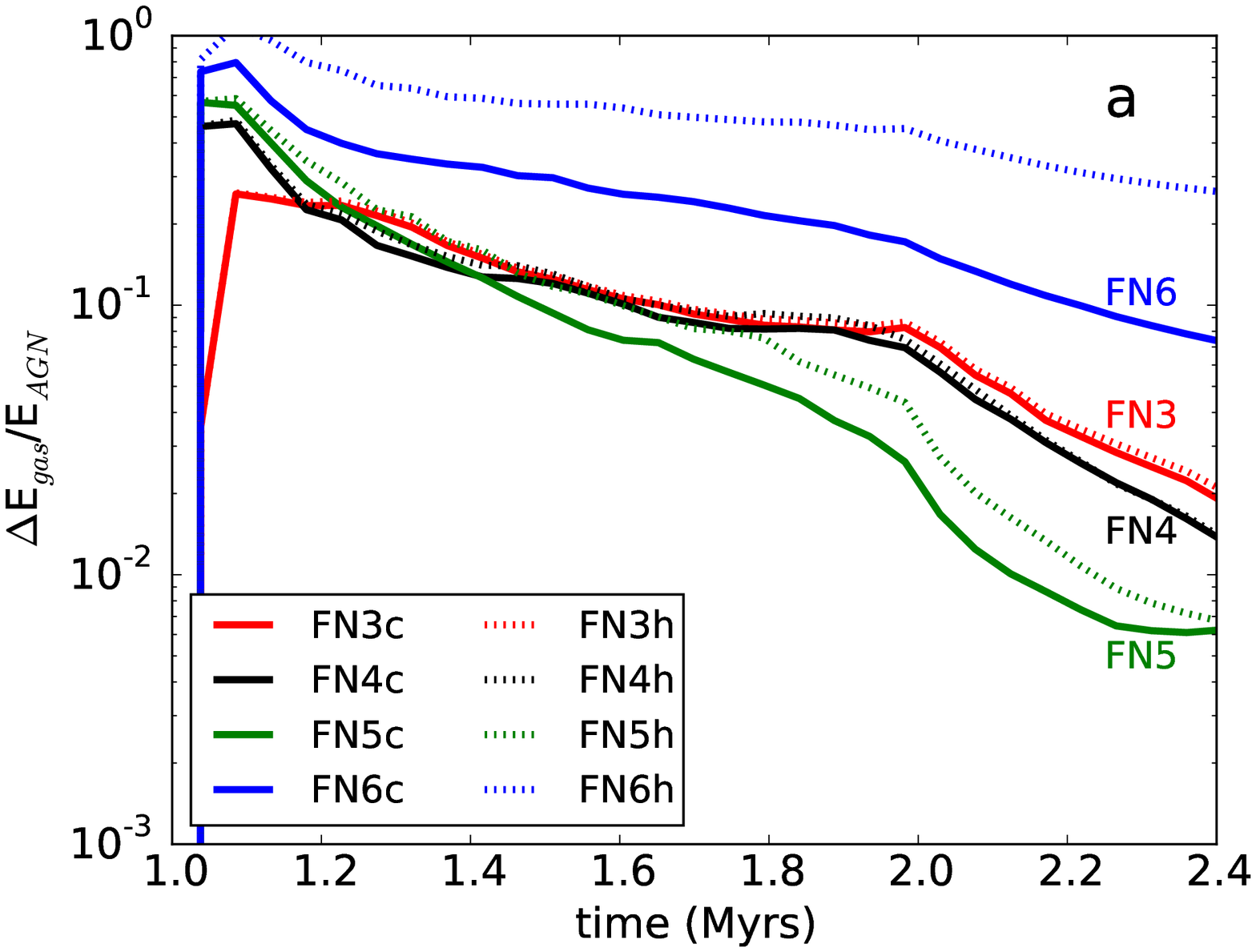,width=0.5\textwidth,angle=0}
\psfig{file=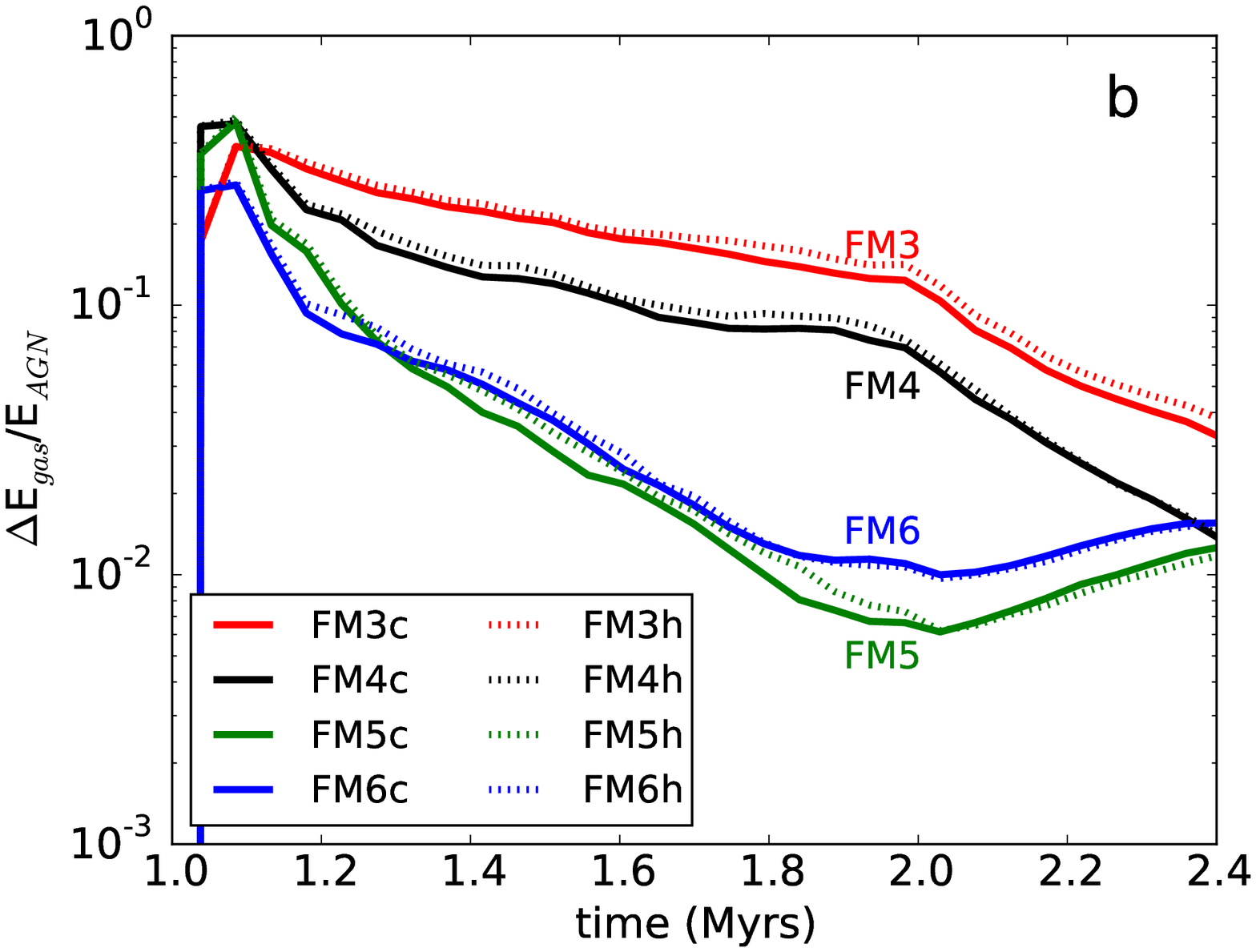,width=0.5\textwidth,angle=0}
\caption{Evolution of the ratio of total gas energy, $E_{\rm gas}$, to total
  AGN input energy, $E_{\rm AGN}$, from the beginning of the AGN outburst
  between 1 Myr and 2 Myr. The top panel (a) shows the FN3 (red), FN4 (black),
  FN5 (green) and FN6 (blue) runs.  In this panel there is no apparent trend
  with resolution. In both panels the solid and dotted lines correspond to
  runs with and without IC cooling respectively. The bottom panel (b) shows
  the FM3 (red), FM4 (black), FM5 (green) and FM6 (blue) runs. There is an
  evident trend that lower resolution runs retain more energy than high
  resolution runs.} 
\label{egas_equas}
\end{figure}

The mode of feedback energy delivery to the ambient gas strongly affects the
subsequent  evolution of the system. This can be seen in  figure
\ref{egas_equas} which shows the time evolution of the ratio of the change in
gas internal and kinetic energy to the energy injected by the AGN as a
function of time. As in figure \ref{Tmax}, in the top panel (a) the feedback
energy is injected into a fixed number of SPH particles ($\sim 100$) during
each time step whereas in the bottom panel (b) the feedback energy is injected
into a fixed mass ($\sim 1.6\times 10^{7} M_{\odot}$) of gas during each time
step. The change in total gas energy at time $t$, $\Delta E_{\rm gas}(t)$, is
given as
\begin{equation}
 \Delta E_{\rm gas}(t)=E_{\rm gas}(t) - E_{\rm gas}(1 {\rm Myr})
\end{equation} 
where 
\begin{equation}
E_{\rm gas}(t)=\sum\limits_{i}\left({\frac{1}{2}m_{\rm i}v_{\rm i}^{2} +
  \frac{3}{2}\frac{k_{\rm B}T_{\rm i}}{\mu m_{\rm P}}m_{\rm i}} \right)
\end{equation} 
is the sum of the kinetic and internal energy of all of the SPH particles at
time $t$. The total AGN input energy, $E_{\rm AGN}$ is given by
\begin{equation} 
E_{\rm AGN}=\epsilon_{\rm f}L_{\rm AGN} t_{\rm act} 
\end{equation} 
where $t_{\rm act}$ is the time for which the AGN has been active. Focusing on
the FN series of runs first (top panel, a), we find no clear trend in how much
feedback energy is retained by the gas. The two low resolution cases, FN3 and
FN4, are rather similar; then the higher resolution case, FN5, retains less
energy than that, but the highest resolution case, FN6, retains {\it much
  more} energy than the low resolution cases. We believe that this is due to
competition between two somewhat oppositely directed numerical artefacts, one
due to the over cooling problem close to the AGN and the other due to poor
sampling of the ambient gas farther out.  As the resolution is improved, one
is able to heat a lower mass of gas and hence heat the gas to higher
temperatures, resulting in longer cooling times. However at higher resolution
one can resolve high density material which has a shorter cooling time than
the lower density material found in the vicinity of the SMBH in lower
resolution simulations. Given the competition between the processes
contributing to the gas cooling rate there is not necessarily a clear trend in
feedback energy conservation with resolution.

The fixed mass runs (FM, the bottom panel, b, of the figure) give a more
consistent picture, with FM5 and FM6 yielding very similar results, perhaps
indicating that a degree of numerical convergence is taking place. Here we see
that at higher resolution less feedback energy is retained in the ambient gas
of the galaxy, presumably because higher density clumps are better resolved at
higher resolutions and they lead to more energy being lost to radiation. These
results show that simulations in which feedback is spread around a fixed mass
of ambient gas should be preferred for numerical reasons to those where the
number of AGN neighbours is fixed instead. However it should be noted that
although injecting feedback energy into a fixed mass of particles provides a
degree of numerical convergence, it is not necessarily convergence towards the
physically correct result. A further potentially confounding factor is the
energy imparted by the momentum of the AGN wind. This is not included in the
models presented in this paper, however a purely momentum-driven wind should
have a kinetic energy rate $\dot{E}_{\rm mom} \simeq 2 \sigma / (\eta c)
\dot{E}_{\rm wind} \simeq 0.01 \dot{E}_{\rm wind}$. In the high-resolution
models, this energy is comparable to the energy retained by the gas and may
further complicate gas behaviour. However, a more detailed investigation of
the effects of different AGN feedback prescriptions is beyond the scope of
this paper. 
  
\subsubsection{Gas ejection efficiency}

It is believed that the most important effect of AGN feedback onto their host
galaxies is to remove gas from the host galaxy and thus quench star
formation. In this section we show that the ability of an AGN to clear the gas
out of the host is greatly affected by numerical resolution at least for the
initial conditions and parameter space investigated in this paper.

\begin{figure} 
\psfig{file=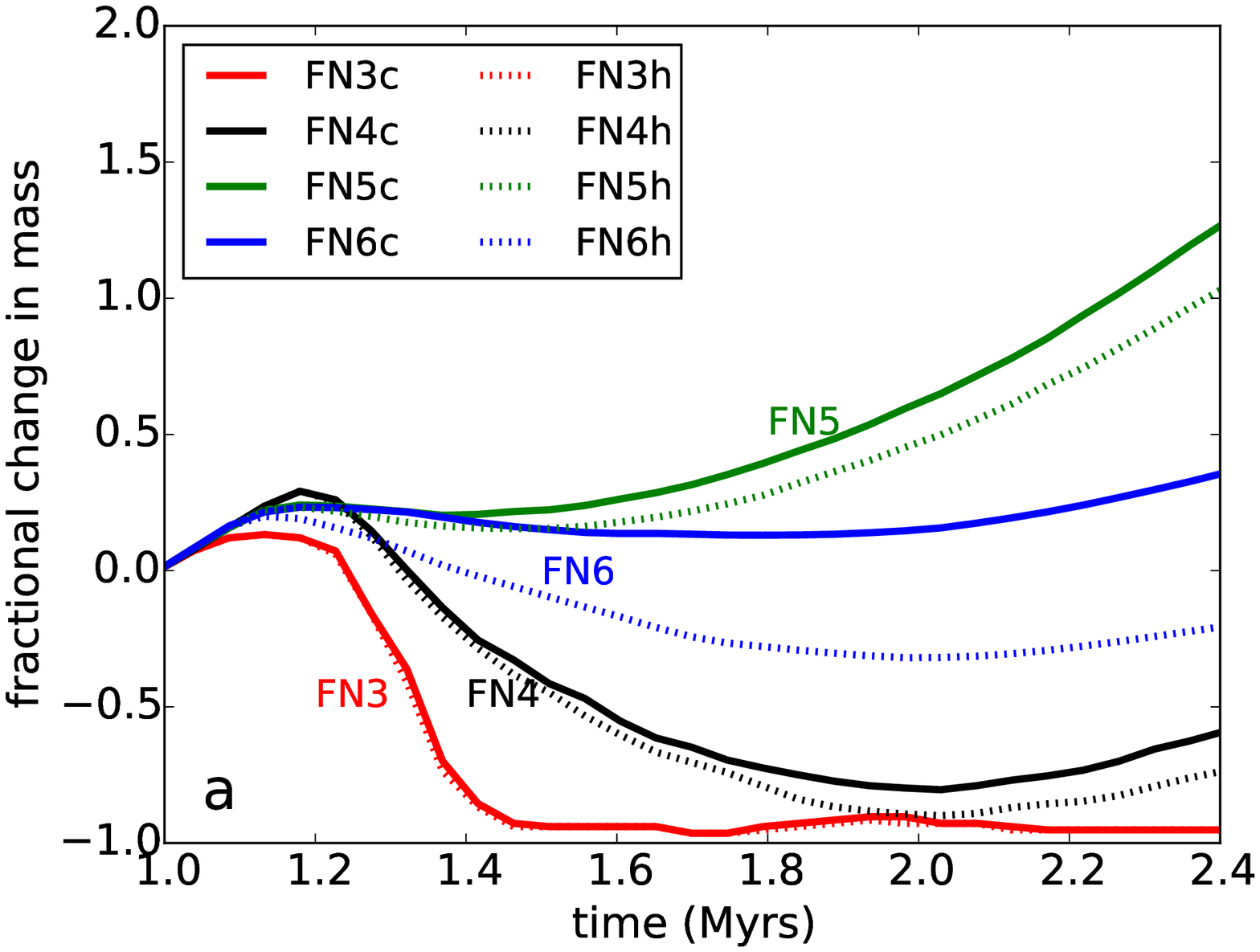,width=0.5\textwidth,angle=0}
\psfig{file=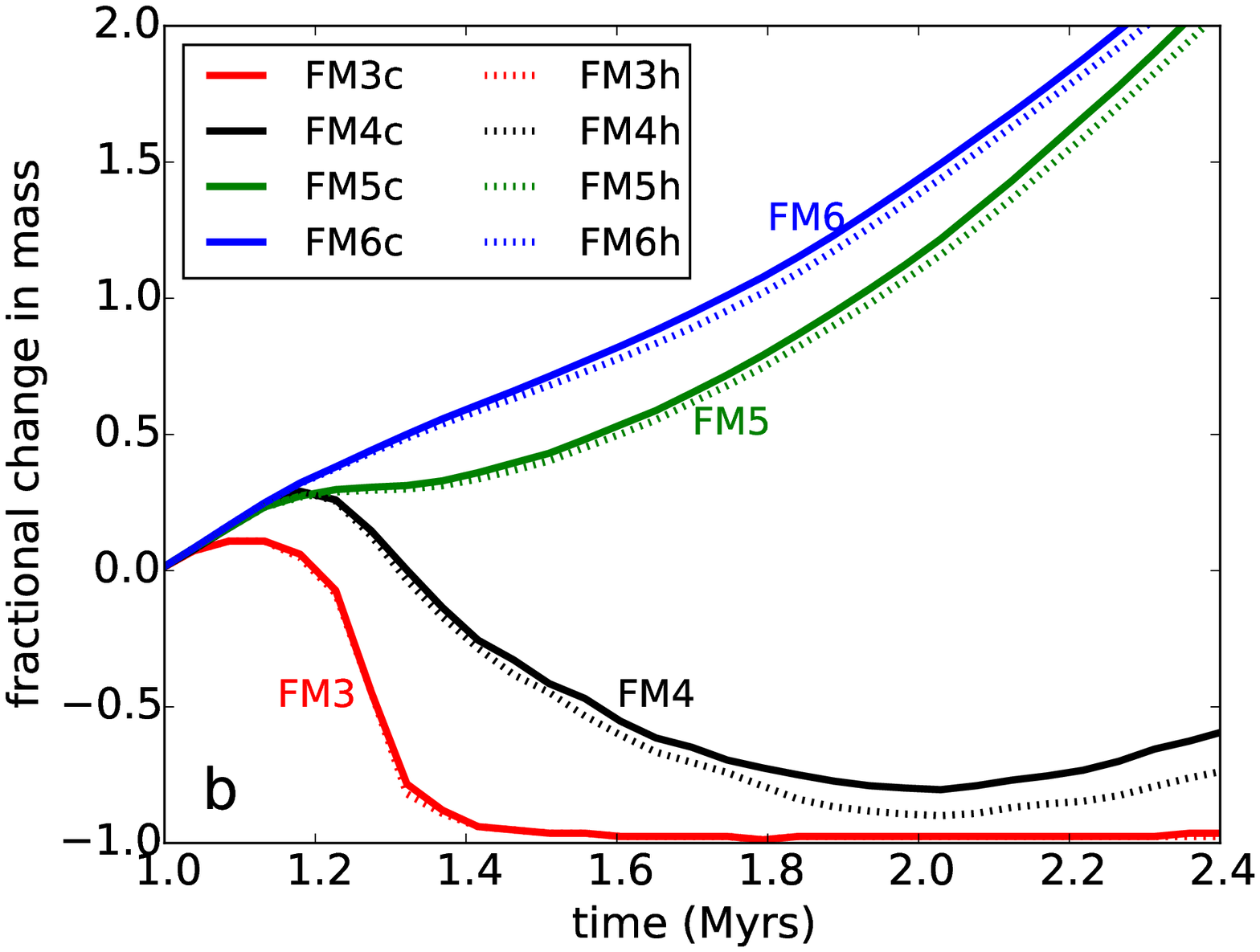,width=0.5\textwidth,angle=0}
\caption{Time evolution of the fractional mass change within the central 200
  pc of the system. The top panel (a) shows the FN3 (red), FN4 (black), FN5
  (green) and FN6 (blue) runs. As in figure \ref{egas_equas} there is no
  apparent trend with resolution, however, like in the FM runs, the FN3 and
  FN4 runs are the most efficient at removing gas. The bottom panel (b) shows
  the FM3 (red), FM4 (black), FM5 (green) and FM6 (blue) runs. As the
  resolution is degraded the feedback becomes more efficient at clearing gas
  from the central regions. In both panels the solid and dotted lines
  correspond to runs with and without IC cooling respectively. } 
 \label{menc} 
\end{figure}

To quantify the AGN feedback impact on the host, we first consider the
evolution of the total baryonic mass enclosed within 200pc of the host's
centre. We define the fractional change in baryonic mass as
\begin{equation}
\frac{\Delta M_{< 200{\rm pc}}(t)}{M_{< 200{\rm pc}}(1 {\rm Myr})}=\frac{M_{<
    200{\rm pc}}(t)-M_{< 200{\rm pc}}(1 {\rm Myr})}{M_{< 200{\rm pc}}(1 {\rm
    Myr})}
\end{equation}
where $M_{< 200{\rm pc}}(t)$ is the total baryonic mass within $200$ pc,
including gas accreted onto the black hole but not the initial black hole mass
( $10^{8}\msun$) itself.  Figure \ref{menc} shows the time evolution of
${\Delta M_{< 200{\rm pc}}(t)}/{M_{< 200{\rm pc}}(1 {\rm Myr})}$ for
simulations with $10^{3}$, $10^{4}$, $10^{5}$ and $10^{6}$ particles shown in
black, blue, red and green, respectively. The fixed mass (FM, bottom panel, b)
runs show a trend with resolution such that feedback becomes less effective at
clearing gas out with improved resolution. The FM3 run which retains the
largest fraction of energy is the most effective at clearing gas out, whilst
the FM5 and FM6 runs, which lose the most energy, cannot prevent the continual
infall of gas. However, if we consider the FN runs (top panel, a) there is no
clear trend. Even though the FN6 run retains substantially more energy than
the FN3 run, it is far less effective at clearing gas out. This can be
attributed to being able to resolve structure in the ISM. As we improve
resolution, the hot gas can escape through paths of least resistance leaving
higher density clumps behind. 
 
\begin{figure} 
\psfig{file=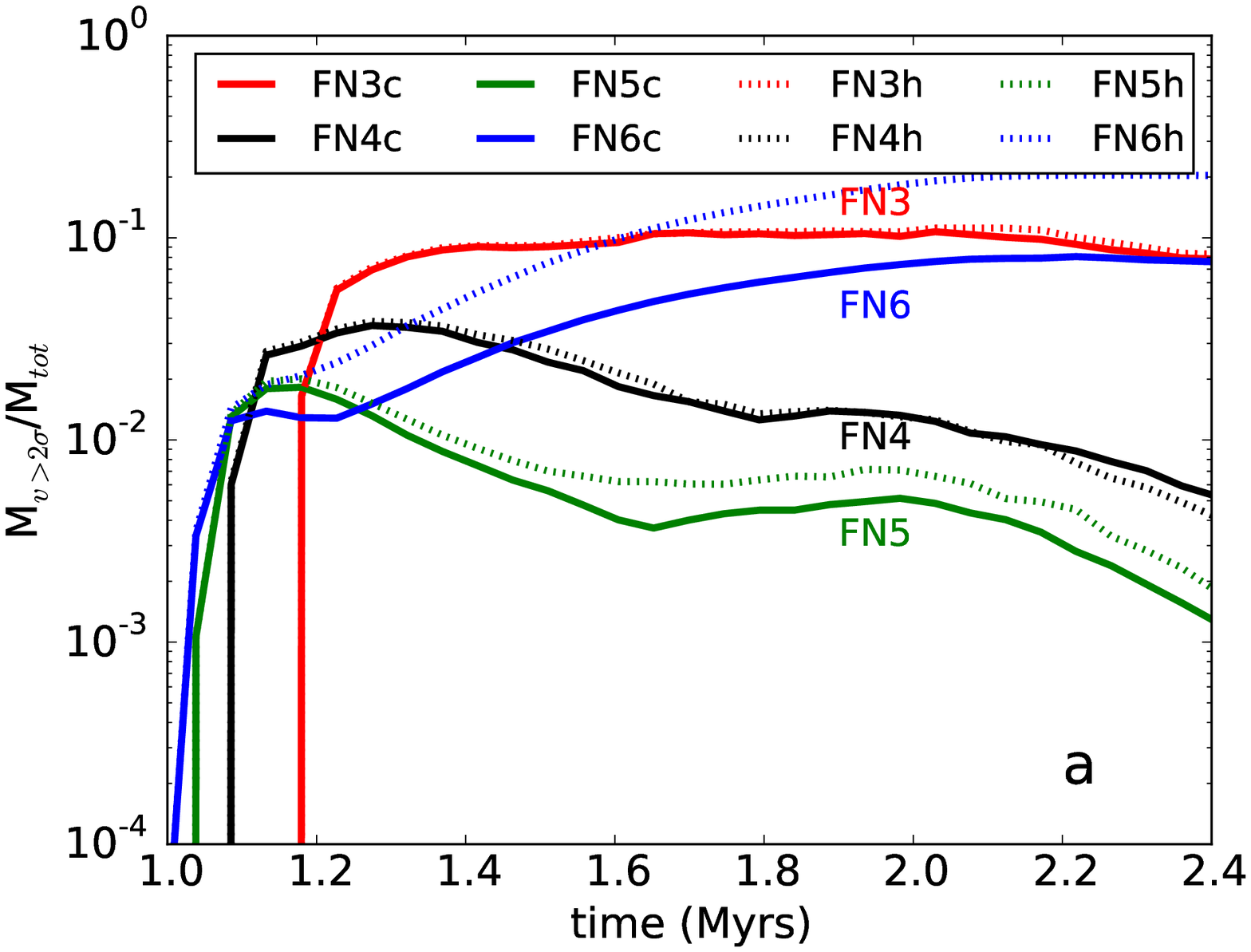,width=0.5\textwidth,angle=0}
\psfig{file=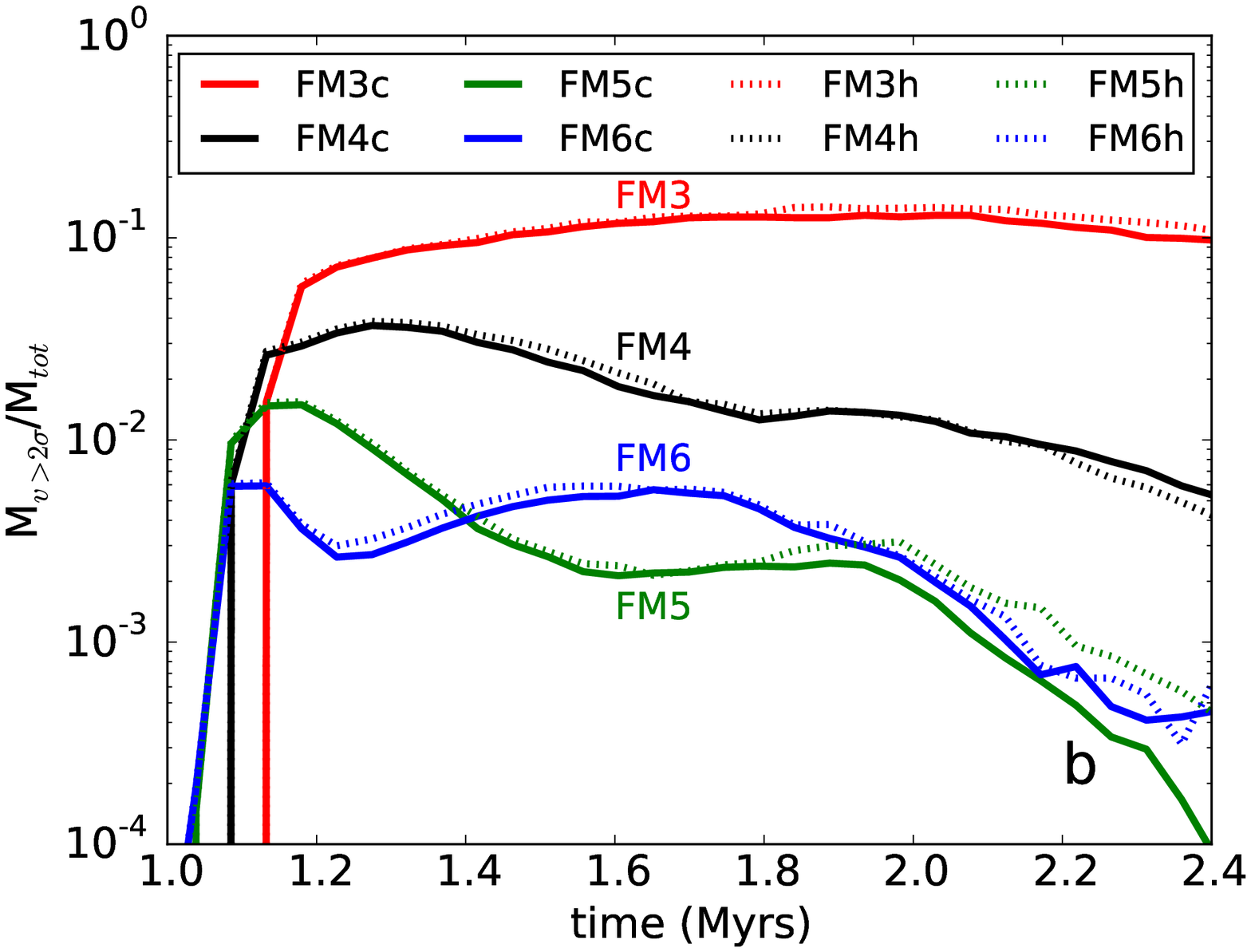,width=0.5\textwidth,angle=0}
\caption{Time evolution of the fraction of gas with radial velocity greater
  than $2\sigma$. The top panel (a) shows the FN3 (red), FN4 (black), FN5
  (green) and FN6 (blue) runs. Again, as in previous figures there is no
  apparent trend with resolution with both the FN3 and FN6 runs producing
  outflows containing similar fractions of gas. The bottom panel (b) shows the
  FM3 (red), FM4 (black), FM5 (green) and FM6 (blue) runs. As in figure
  \ref{menc} there is an apparent trend with resolution, with a greater
  fraction of gas being out-flowing at a high velocity in the lower resolution
  runs than in the high resolution runs.  In both panels the solid and dotted
  lines correspond to runs with and without IC cooling respectively.} 
 \label{eff} \end{figure}

Finally we attempt to further quantify the ability of an AGN to clear gas out
by calculating the fraction of gas with radial velocity greater than
$2\sigma$. This is shown in figure \ref{eff}, with the same colour scheme as
previous plots. For the FM runs (bottom panel, b) there is a clear trend that
as the resolution is improved the fraction of gas out-flowing at a high
velocity becomes smaller. However for the FN runs (top panel, a), this trend
breaks down for the highest resolution run, FN6. This suggests that the
cooling as well as the ability to resolve structure plays an important role in
determining how effectively gas can be blown out of the galaxy. As discussed
when considering figures \ref{cross_T} and \ref{Tmax}  with respect to the FN5
and FN6 runs, the gas in the FN6 runs is heated to higher temperatures than
that in the FN5 runs. The cooling in the FN6 runs is less efficient and thus
the hot feedback bubble conserves more energy than in the FN5 runs and is able
to drive more powerful outflows.  

\section{Discussion and summary}

 We have studied the effect of AGN feedback on a multiphase interstellar
 medium and how such feedback is affected by numerical resolution. The
 resolution has two competing effects on the results. Unsurprisingly, the
 density structure is better resolved at higher resolutions, so that there is
 more hot low-density and also cold high-density gas than in low resolution
 simulations. Low resolution simulations thus tend to expel all the gas from
 the centres of host galaxies, whereas higher resolution simulations do not;
 they also show dense clumps that are more resilient to feedback and may
 continue to fall inward while most of the gas is driven out. On the other
 hand, the over-cooling problem may affect the gas in the immediate vicinity
 of the SMBH, actually reducing feedback efficiency in low resolution
 simulations. We now discuss these opposing effects in detail.

\subsection{Resolving the multiphase ISM and outflow properties}

 The inferred properties of outflows in our simulations strongly depend upon
 the resolution and how the feedback energy is coupled to the ISM. At low
 resolution the feedback sweeps up {\it all} the material in SMBH vicinity
 into an outflow with a modest velocity ($\sim 1000$ km s$^{-1}$). In
 contrast, at high resolutions only some of the neighbouring gas is launched
 into the outflow. The outflows can reach higher radial velocities (up to
 $\sim 5000$ km s$^{-1}$); however, cold dense filaments may continue to fall
 in and feed the SMBH. Higher resolution SPH simulations thus lead to a much
 more varied outcome for the gas in the galaxy in every measurable quantity,
 e.g., density, temperature and velocity. This is potentially very important
 for SMBH growth since even a tiny mass of ambient gas is sufficient to
 increase the SMBH mass significantly.   

 In the same vein, cosmological simulations often have to invoke high {\it
   stellar} feedback efficiencies in order to produce observed galactic winds
 \citep{Schaye14}. One possible reason for this, in addition to the
 over-cooling problem, is that low resolution inhibits low density channels
 through which outflows can escape to reach galactic scales. Alternatively, in
 some cosmological and galaxy-scale simulations, such winds are
 hydrodynamically decoupled from the ISM so that they can freely stream to
 galactic scales \citep{Springel03, Oppenheimer06, Oppenheimer10,
   Puchwein13}. While such {\it ad-hoc} prescriptions allow one to produce
 realistic outflows at large radii, one loses any information regarding the
 direct interaction of feedback with the ISM. It is therefore clear that
 cosmological simulations are unable to provide detailed insights into the
 feedback mechanisms themselves and the best that one can hope for is that
 their effects on resolvable scales are modelled correctly.

\subsection{Cooling of the feedback bubble}

It is evident that the ability of the feedback bubble to conserve its energy
has a significant impact on how efficient it is in destroying the host
galaxy. This can depend intimately both upon the temperature to which gas is
heated directly by the AGN and the processes through which the feedback heated
bubble cools. Considering the first point, we have shown that there are stark
differences in the properties of the feedback depending upon the mass of the
gas heated by AGN feedback. This leads us to the question, what temperature is
correct?  Analytical theory suggests that wind shocks have temperatures of
$10^{10}-10^{11}$ K, however, as illustrated by figure \ref{Tmax}, in order to
reach such high gas temperatures in a cosmological simulation, a gas mass of
$\simlt 1 $ SPH particle would need to be heated. Therefore, in such
simulations, gas is typically heated to lower temperatures, which could
potentially result in incorrect energy conservation in the hot bubble. Such
problems may be mitigated to some degree by {\it tuning} the efficiency of the
feedback or artificially turning off radiative cooling in order to match
observations. Whilst such procedures can lead to correct large scale
properties of the galaxies, e.g.  correct stellar masses and the $M_{\rm
  BH}-\sigma$ relation, one clearly loses predictive power if observations
need to be used to calibrate the models.

With regard to our second point, i.e. the relevance of different cooling
mechanisms, the inclusion or absence of IC processes can have a big effect.
Comparing the FN6c and FN6h runs, it is clear from figures \ref{egas_equas},
\ref{menc} and \ref{eff} that the hot bubble retains more energy and clears
out more gas when IC cooling is neglected. It is therefore important to
understand which scenario is more physically motivated. \citet{FQ12a} have
shown that given the high temperature and low density properties of shocked
outflows, the electrons and ions are thermally decoupled. The electron cooling
timescale is shorter than the electron-ion thermalisation timescale and
therefore it is the latter that determines the cooling rate of the gas,
with IC cooling becoming ineffective. This suggests that the runs in which IC
cooling is neglected are more physically motivated. However, even if the mass
and temperature of the feedback bubble matches those expected from analytical
theory, the simulated feedback bubble likely has a higher density than the
actual shocked wind bubble would, because we are directly heating ISM
gas. Therefore the cooling of the hot bubble and its interaction with the
ambient ISM may still be incorrectly modelled. Thus we conclude that direct
physically self-consistent modelling of AGN feedback heating and cooling on
small scales is still beyond the reach of modern numerical capabilities.

\subsection{Star formation during an AGN outburst}

Figures \ref{rho_dist_comp_ngb} and \ref {rho_dist_comp_mass} clearly show
that gas can be compressed to high densities by an AGN outflow. The presence
of dense gas can result in additional star formation, which both
quantitatively and qualitatively changes the properties of the AGN host
galaxy. Similar aspects of AGN-triggered star formation have already been
explored by \citet{NZ12} and \citet{ZubovasEtal13b}, who found that
significant star formation can occur both in the cooling out-flowing medium
and in the compressed disc of the host galaxy. Our results show that any
density contrasts can be enhanced by AGN outflows.  

It is important that one needs simulations with sufficient resolution to
recover the compression effect in numerical simulations (Figures
\ref{rho_dist_comp_ngb} and \ref {rho_dist_comp_mass}). Large-scale
simulations with low numerical resolution typically miss this effect and hence
over-predict the negative (gas removal) effect of AGN feedback. Even in
high-resolution simulations, star formation in dense clumps is difficult to
track during the AGN outflow if one employs a heating-cooling prescription
such as that of \citet{sazonovetal05}, which includes Compton heating. This
prescription assumes that gas is optically thin, which is a good approximation
for the low density ISM, but not for the dense clumps. As a result, the clump
temperature, in general, stays too high (i.e. above the temperature set by
equation \ref{T_floor}) and fragmentation is slower than it would be with a
proper radiative transfer treatment.  

The lack of AGN-triggered star formation in low-resolution simulations of
galaxy evolution presents two challenges: quantitative and qualitative. The
quantitative challenge is the issue of reconciling the star formation
histories of simulated and observed galaxies. If one channel of triggered star
formation is missed in simulations, the other star formation channels have to
be proportionately enhanced (for example, by adopting higher star formation
efficiencies or lower density thresholds) in order to reproduce the galaxy
stellar mass functions of present-day galaxies. The qualitative challenge is
arguably more important: AGN outflows create dense star-forming gas where
there was none, i.e. affect the location of star-forming regions in the
galaxy. This process directly affects the morphology of the starburst and the
dynamics of new-born stars \citep{NZ12, ZubovasEtal13a}. Both of these effects
are missed in low-resolution simulations; however, they can be used as strong
indicators of positive AGN feedback.  

One region where AGN-induced star formation may be particularly important is
galaxy centres. These regions typically contain dense gas discs or rings
\citep[e.g.][]{BoekerEtAl08}, which often show clumpy structures and embedded
young star clusters. It is generally accepted that star formation in these
regions is induced by shocks caused by matter in-falling via galactic bars
from larger radii. However, the presence of young clusters and the lack of
azimuthal age gradient in some of these systems complicate this picture
\citep{BoekerEtAl08}. Another trigger of star formation in these systems could
be AGN outflows. As these outflows expand perpendicularly to the disc plane
due to lower density \citep{ZN14}, they significantly compress the gas in the
midplane; in addition, ram pressure of the AGN wind pushes the disc gas into a
narrower ring, which is more prone to gravitational instability (Zubovas,
submitted). In this way, the density contrast between the disc and its
surroundings is also enhanced, much like the density contrast between
different regions in the simulations presented in this paper. 

\subsection{Black hole growth and the $M_{\rm BH}-\sigma$ relation}

 The general trend in the simulations we present is that at higher resolution
 less gas is cleared out than at lower resolutions. In low resolution runs the
 outflow sweeps up everything in its path, creating a sharp cut-off radius
 between out-flowing material and in-flowing material. However in high
 resolution runs the outflow only sweeps up low density material, whilst high
 density material can continue to flow inward. This could lead to very
 different feeding cycles for the black hole. The supply of material to the
 black hole is completely cut off and cleared to large radii in the low
 resolution runs, whereas in the high resolution runs clumps and filaments can
 remain in-flowing at small radii and thus continuously feed the black
 hole. This sets up a scenario in which feedback is {\it ``all or nothing''}
 at low resolution but more diluted at high resolution, with feeding becoming
 interminable up to the point that the gas can form stars.  

In the high resolution scenario, the high density clumps will only be acted
upon by the momentum of the AGN wind \citep[BNH14,][]{Nayakshin14} with the
energy escaping through low density channels. By requiring the ram pressure of
the AGN outflow to exceed the gravitational force of the bulge acting on all
of the clumps along the line of sight from a SMBH and setting a maximum
threshold density for the clumps (assumed to be the density at which they
undergo star formation) \cite{Nayakshin14} finds a critical black hole mass in
order to clear out the cold gas of
\begin{equation}
M_{\rm crit}\sim 2.2\times 10^{8}\msun\sigma_{200}^{4}
\end{equation}
comparable to the observed $M_{\rm BH}-\sigma$ relations
\cite[e.g.,][]{KormendyHo13}. 

\subsection{Comparison with previous work}

 Recent work \citep[BNH14]{wagner13} has shown that the structure of the ISM
 can impact upon the ability of an AGN to clear out gas and hence quench star
 formation. BNH14 present high-resolution simulations of an UFO impacting upon an inhomogeneous,
 turbulent medium and find new processes such as energy leakage and
 separation of energy and mass flows within the ISM. The
 shocked outflows escape via paths of least resistance, leaving the high density gas, which is difficult to expel, largely intact. Such
 processes have previously been missed in analytical models and cosmological
 simulations mainly because the multiphase nature of the ISM is not
 resolved and has to be implemented as a sub-resolution model. For example
 \citet{Springel03} include a sub-grid multiphase model for star formation
 while \citet{Murante2010} include a non-equilibrium model that includes the
 three ISM phases for each SPH particle. Unfortunately such methods mean that
 the intricate structure one expects is washed out due to low resolution.   

This work builds on that of BNH14 by implementing a continuous, Eddington
limited feedback outburst, rather than a single hot bubble. Furthermore, we
have explored the role of IC cooling against the AGN radiation field, which
has been highlighted in the literature as a key feature in understanding the
impact of UFOs \citep{FQ12a}. This work adds to the growing body of work
\citep[e.g.][BNH14]{wagner12, wagner13,costa14,gabor14,ZN14} highlighting that
the AGN {\it environment} can be just as important as the AGN feedback
mechanism itself when modelling galaxy evolution. In BNH14, the main aim was
to understand the physics of the interaction of an outflow with the multiphase
ISM, however in this work we have focused more on how resolution can affect
this interaction.  

The role of the ISM and its impact on AGN feedback has been studied by a
number of authors both for feedback in the form of jets
\citep[e.g.][]{wagner12} and UFOs
\citep[e.g.][BNH14]{wagner13}. \cite{wagner12} present high resolution
simulations of jet feedback in a clumpy ISM. They found that if the volume
filling factor of the clouds is less than $0.1$ then the hot feedback bubble
can expand as in the energy driven limit. Clouds smaller than $\sim 25$ pc are
destroyed and dispersed, leading them to argue that feedback prescriptions in
cosmological simulations should provide a good description of this regime as a
source of negative feedback. However if clouds are larger than $\sim 25$ pc
they are more resilient to the feedback. In agreement with this work they find
that the clouds can be compressed, potentially triggering star formation. Such
behaviour is missed in cosmological simulations. Whilst \cite{wagner12}
suggest a physical setup in which feedback prescriptions in cosmological
simulations may produce correct results, we provide a direct comparison of the
nature of feedback when simulated at low resolution (similar to cosmological
simulations) and up to three orders of magnitude higher resolution. We have
found that across such a resolution range there are marked differences in the
evolution of the feedback, caused by a combination of effects including the
ability to resolve structure and the thermal and physical properties of the
hot feedback bubble.  

Further work on scales simulating whole galaxies has shown that large scale
structure such as a disk \citep{gabor14} or filaments \citep{costa14} can also
reduce the ability of AGN feedback to remove gas from the galaxy. Such
structure should be resolved in cosmological simulations, however resolution
effects can still impact upon the properties of the feedback. As we have
shown, the temperature to which gas is directly heated by feedback as well as
its density can affect the cooling and thus the efficiency of the
feedback. Given such large differences between the physical properties of hot
feedback bubbles in cosmological simulations and those expected in reality we
should pose the question when, if ever, such processes can be included in
these simulations. Following a Moore's law approach the number of particles
used in cosmological N-body simulations approximately doubles every 16
months. This would suggest that an increase in the mass resolution by 3 orders
of magnitude could be achieved in $\sim 13$ years. However, given the fact
that algorithms typically scale worse than O(N), and also considering that the
silicon chip capacity is limited, this is an extremely optimistic estimate. It
is therefore likely that such an improvement in resolution would take much
longer to achieve and depends upon the efficiency with which simulators and
programmers can harness the power of parallel processing and other
technological advances. 

\subsection{Implications for cosmological simulations}

A caveat to the results presented in this paper is that our simulations
  do not include self-regulation of the AGN feedback, which plays an important
  role in galaxy evolution. Instead our simulations only model a single, 1 Myr
  long, Eddington limited AGN feedback event which is not linked to the gas
  content of the host galaxy. It may therefore be argued that our
results on the numerical artefacts in AGN feedback efficiency do not have
direct implications for cosmological simulations. In such simulations, the
system will undergo multiple feedback events over cosmological timescales. The
rate at which a black hole injects energy into the host galaxy ISM is
  coupled to the gas accretion rate $\dot{m}_{\rm accr}$ through the equation
\begin{equation}
\dot{E}=\epsilon_{\rm BH}(M_{\rm BH}, \dot{m}_{\rm accr})\dot{m}_{\rm
  accr}c^{2}
  \label{edot_accr}
\end{equation}
where $\epsilon_{\rm BH}(M_{\rm BH}, \dot{m}_{\rm accr})$ is as defined in the
introduction and can be a function of $M_{\rm BH}$ and $\dot{m}_{\rm accr}$.
For example \citet{DavisLaor11} determined $\epsilon_{\rm r}$ in 80
  quasars by using their bolometric luminosities and absolute accretion rates,
  calculated using thin accretion disk model spectral fits, finding a scaling
  with $M_{\rm BH}$ such that $\epsilon_{\rm r}=0.089 M_{8}^{0.52}$,
  where $M_{8}$ is $M_{\rm BH}$ in units of $10^{8}$ M$_{\odot}$.
Often, however, $\epsilon_{\rm BH}(M_{\rm BH}, \dot{m}_{\rm accr})$ is set to
be a constant for simplicity.

The coupling of the AGN feedback to the gas content of the galaxy through
equation \ref{edot_accr} leads to self-regulation of the SMBH growth and
feedback resulting in the {\it correct} $\dot{E}$ such that the feedback
driven outflows balance mass inflow. This, therefore, does not uniquely
establish $M_{\rm BH}$ but rather the product $\epsilon_{\rm BH} M_{\rm BH}$
(because $\dot{m}_{\rm accr}$ is usually limited to the Eddington accretion
rate). To reproduce the observed black hole correlations, one {\em fixes} the
value of $\epsilon_{\rm BH}$ \citep[e.g.,][]{BoothSchaye09, Booth2010,
  Schaye14}. Further, provided that $\epsilon_{\rm BH}$ is set to a value
within a suitable range, the observed SMBH scaling relations can be reproduced
despite large differences in resolution and sub-grid prescriptions
\citep[e.g.,][]{DiMatteo05, SpringelEtal05, Sijacki07, BoothSchaye09,
  Schaye10, Schaye14}, although some fine tuning may be required (see
discussion in the introduction). A key element of self-regulation is that the
physical properties of the galaxies, such as the stellar mass
\citep[e.g.][]{DiMatteo05, SpringelEtal05, Sijacki07, BoothSchaye09,
  Booth2010} or AGN driven outflow rates \citep{Schaye14} do {\it not} depend
upon the chosen value of $\epsilon_{\rm BH}$. The result of this is that any
dependencies that $\epsilon_{\rm BH}$ has on resolution whould not effect the
global properties of the galaxy due to self-regulation, although may lead to
changes in the SMBH mass.

As an alternative to tuning efficiencies, a number of authors have attempted a
more physically constrained approach to AGN feedback, in which $\epsilon_{\rm
  BH}(M_{\rm BH}, \dot{m}_{\rm accr})$ is not a free parameter. For example,
we here followed the model of \cite{king05} in setting $\epsilon_{\rm f} =
0.05$. Additionally, there is growing evidence that AGN should undergo
separate {\it quasar} and {\it radio} modes of feedback
\citep[e.g.,][]{Churazov2005, Heinz2005, Croton06, Ishibashi2014} depending on
the Eddington ratio, $\dot{m}_{\rm accr}/\dot{M}_{\rm Edd}$, each with
differing values for $\epsilon_{\rm f}$. It has also been suggested that
$\epsilon_{\rm r}$ depends upon $\dot{m}_{\rm accr}$
\citep[e.g.,][]{Narayan95, Mahadevan1997, CiottiOstriker01, Ciotti2009}, with
some cosmological simulations already attempting to include this additional
physics \citep[e.g.,][]{Sijacki07, Sijacki2014,Vogelsberger2014}. Further, an
effect that is not typically taken into account in galaxy formation
simulations is that of the black hole spin, which can lead to variations in
the radiative efficiency in the range of $0.055 < \epsilon_{\rm r} < 0.42$.

We believe that the future of the field is in these more physically motivated
approaches which would hopefully provide predictions constraining the physics
of black hole growth and feedback. Despite our simulations lacking
  self-regulated AGN feedback our results are still important for such
an approach. As discussed in the introduction some fine tuning of sub-grid
feedback parameters can still be necessary when changing resolution
\citep[e.g.,][]{Schaye14, Crain2015}, sub-grid ISM models
\citep[e.g.,][]{BoothSchaye09} or cooling prescriptions
\citep[e.g.,][]{Sijacki2014}. When comparing results of simulations with
different AGN feedback physics to the observations, one must be acutely aware
of numerical artefacts that may skew the interpretation of such comparisons. 

Further we note the potential dependence that $\epsilon_{\rm BH}$ has with the
spatial resolution of the SMBH surroundings. If a simulation probes this
parameter on sub-pc scales, then the factor will determine the efficiency of
BH wind production; on pc scales, the factor tells us something about the
coupling between the wind and the surrounding ISM (perhaps about the
clumpiness of the ISM); on scales of tens or hundreds of pc, the factor also
encompasses the thermal effects (mostly cooling, but perhaps also heating by
the AGN radiation field) of the gas surrounding the AGN. Therefore simulations
with different spatial resolution might be probing different processes which
contribute to AGN feedback efficiency. This is an important point to consider
when interpreting constrained values of $\epsilon_{\rm BH}$.

\section{Summary}

 In this paper we have studied the effect of an Eddington-limited AGN outburst
 on a multiphase  turbulent ISM, with particular focus on the effects of
 numerical resolution.  In general, at higher numerical resolution, more dense
 clumps and also voids through which the feedback can escape are found. This
 reduces the efficiency with which AGN feedback clears out the host's gas. At
 low resolution this behaviour is lost as the feedback sweeps up essentially
 all the gas in its path. Additionally, depending on uncertain physical detail
 of the radiative cooling function for the gas heated by AGN feedback,
 numerical resolution also affects the amount of AGN feedback energy lost to
 radiation, and it is not possible to say whether it will increase or decrease
 the feedback efficiency in a general case. It is therefore plausible that
 resolution dependent effects alter the efficiency of AGN feedback in such a
 way that it is difficult to attach solid physical meaning to constrained
 values of $\epsilon_{\rm BH}=\epsilon_{\rm f}\epsilon_{\rm r}$. We also
   note that although over cosmological timescales self-regulation results in
   consistent galaxy properties and outflow rates irrespective of the chosen
   feedback efficiency, our simulations illustrate certain physical processes,
   such as energy leakage through a clumpy ISM, that can only be modelled at
   sufficiently high resolution.  Finally, in agreement with \cite{Schaye14},
 we therefore suggest caution when trying to ``invert'' the results of
 cosmological simulations (usually tuned to fit observations) to learn about
 certain physical aspects of AGN feeding and feedback.

\section*{Acknowledgments} 
We would like to thank the anonymous referee for detailed comments which have helped to improve this paper. We acknowledge an STFC grant. MAB is funded
by a STFC research studentship. KZ is funded by the Research Council Lithuania
grant no. MIP-062/2013. We thank Hossam Aly for useful discussions, and Justin
Read for the use of SPHS. This research used the DiRAC Complexity system,
operated by the University of Leicester IT Services, which forms part of the
STFC DiRAC HPC Facility (www.dirac.ac.uk).  This equipment is funded by BIS
National E-Infrastructure capital grant ST/K000373/1 and STFC DiRAC Operations
grant ST/K0003259/1. DiRAC is part of the UK National
E-Infrastructure. Figures \ref{cross_rho} and \ref{cross_T} were produced
using SPLASH \citep{Price07}.  \bibliographystyle{mnras}
\bibliography{nayakshin} \label{lastpage}
\end{document}